\documentclass[aps,reprint,pre,floatfix,twocolumn,noshowpacs,superscriptaddress]{revtex4-1}

\usepackage{graphicx}
\usepackage{amsmath,amsfonts,mathrsfs}
\usepackage{epsfig,cleveref,nameref}
\usepackage{xcolor}
\usepackage[caption=false, font=footnotesize, justification=RaggedRight]{subfig}
\graphicspath{{./Figs/}}

\usepackage{array}
\newcolumntype{H}{@{}>{\lrbox0}l<{\endlrbox}}
\usepackage{float}
\begin{document}

\title{Going beneath the shoulders of giants: tracking the cumulative knowledge spreading in a comprehensive citation network}
\author{Pietro della Briotta Parolo}
\affiliation{Department of Computer Science, School of Science, Aalto University, Finland} 
\author{Rainer Kujala}
\affiliation{Department of Computer Science, School of Science, Aalto University, Finland} 
\author{Kimmo Kaski}
\affiliation{Department of Computer Science, School of Science, Aalto University, Finland}  
\affiliation{The Alan Turing Institute, British Library, UK} 
\author{Mikko Kivel\"a}
\affiliation{Department of Computer Science, School of Science, Aalto University, Finland} 
\date{\today}

\begin{abstract}
In all of science, the authors of publications depend on the knowledge presented by %
the previous publications. Thus they ``stand on the shoulders of giants'' and there is a flow of knowledge from previous publications to more recent ones.   
The dominating paradigm for tracking this flow of knowledge is to count the number of direct citations, but this neglects the fact that beneath the first layer of citations %
there is a full body of literature.
In this study, we go underneath the "shoulders" by investigating the cumulative knowledge creation process in a citation network of around 35 million publications. In particular, we study stylized models of \textit{persistent influence} and \textit{diffusion} that take into account all the possible chains of citations.
When we study the persistent influence values of publications and their citation counts, we find that the publications related to Nobel Prizes i.e. Nobel papers have higher ranks in terms of persistent influence than that due to citations, and that the most outperforming publications are typically early works leading to 
hot research topics of their time.
The diffusion model %
reveals a significant variation in the rates at which different fields of research share knowledge.
We find that these rates have been increasing systematically for several decades, which can be explained by the increase in the publication volumes. 
Overall, our results suggest that analyzing cumulative knowledge creation on a global scale can be useful in estimating the type and scale of scientific influence of individual publications and entire research areas as well as yielding insights which %
could not be discovered by using only the direct citation counts.
\end{abstract}

\maketitle
\section{Introduction}

Since the seminal work of de Solla Price \cite{deSollaPrice510} quantitative analysis of knowledge spreading through a network of scientific publications has become a matter of great interest. The analysis of bibliometric data not only sheds light on the structure of science and its knowledge accumulation, but also gives us insight into the citation distributions \cite{Wallace2009296,RednerStatistics,Radicchi11112008}, collaboration networks \cite{Newman16012001,Barabasi2002590}, geographical patterns of collaborations and citations \cite{pan2012world,Havemann2006,Jones2008}, and the structural changes that take place at the level of scientific fields \cite{Sinatra2015,Rosvall29012008,10.1371/journal.pone.0010355}. 
Along this line of research, the citations between scientific publications are in the focus of interest and they can encode various meanings between publications \cite{HURT19871}, but perhaps most often they indicate that some knowledge from the cited publication is being used in the citing publication \cite{Bornmann2008}.
Despite all this progress for more than half a century, a core question remains elusive: at the global scale, where is the knowledge going and where is it coming from?

The main research paradigm in the \emph{Science of Science} has been to focus locally on the direct citations between a pair of publications. This thinking is exemplified by the literature on quality measures that are based on direct citations, such as the H-index \cite{10.2307/4152261}, the Journal Impact Factor \cite{Garfield1999}, and a number of others. Even though much attention has been given to the structural limitations of these methods \cite{Penner2013,Adler2009,Alonso2009}, the standard approach to overcome such limitations has been to introduce minor adjustments while still relying on the numbers of direct citations each publication/author/journal receives
\cite{BrasAmorós2011248,doi:10.1007/s11192-006-0090-4,Braun2006,Egghe2006}. 
This local paradigm is in contrast to the structure of science itself because science is a cumulative process where researchers ``stand on the shoulders of giants'', i.e., the results of each researcher are intrinsically based on a massive amount of previous work, not just the publications that are directly cited \cite{Merton1957,Bornmann2008}. Therefore, when one attempts to study the structure and behaviour of scientific knowledge accumulation, it is necessary to look at the whole process and not focusing only on a local area of the system.

In this work, we aim at answering the following question: starting from a publication or group of publications, where and how does its knowledge flow in a citation network if one looks beyond the direct citations? To answer this question, we study all the possible chains of citations, indicated by a citation network, the  publications form. In particular, we introduce two stylized models for studying the flow of knowledge, i.e. the models of \emph{persistent influence} and \emph{diffusion}.
These stylized models are complementary to each other and are computationally tractable even if they are applied to the citation network of all the scientific publications. Here, we use these models to study a comprehensive data set of around 35 million publications from most fields of science covering more than hundred years of making research (see Appendix \ref{sec:data} for details).

When we use millions of individual publications and different groups of them as the sources of information chains, we obtain a good general understanding of the knowledge flows in the citation network. When we compare the relationship between the direct citation counts and the total persistent influence of individual publications, we find that papers associated with Nobel prizes tend to outperform their peers with similar citation counts and publication years. Given that papers related to Nobel prizes are expected to have a profound long-term impact on science, this result validates in part our global approach to model the spreading of scientific knowledge.

There are some previous works looking at the impact of citations beyond the local perspective, but often it was done from a very different starting point as compared to the present work. The spread of information has been studied through the adoption rates \cite{Kim2014}, contagion models \cite{KISS201074}, and diffusion \cite{Gao2012,ASI:ASI23541}, but these studies have either relied on small samples of the full citation network or of the aggregated networks. This is in contrast to our approach to model the processes on a comprehensive network of millions of publications, which allows us to track individual publications and where the processes are independent of any sampling or categorisation of journals and publications. Further, the PageRank-type algorithms have been used to rank publications \cite{Chen20078} and individual scientists \cite{PhysRevE.80.056103} in a set of physics journals. These PageRank-type methods are closely related to diffusion but they rely on random walkers that do not keep track of their origins and destinations, whereas we are interested in how the information from one publication (or collection thereof) is used in other publications.

In addition, it is possible to quantify the spreading of scientific ideas, or memes, between citing and cited publications \cite{PhysRevX.4.041036}. Perhaps the closest work to ours is the study of in-components of individual publications in a citation network of physics papers~\cite{10.1371/journal.pone.0113184}, which has enabled to pinpoint influential, but low-cited, publications of Nobel-Prize winners.

This paper is organised as follows. First we introduce the \textit{persistent influence} model and use it to track the amount of influence the millions of individual publications have on all other publications downstream to them in the citation data we use. After summarising these results, we explore how the papers associated with a Nobel Prize perform in terms of persistent influence, and explore the publications whose rankings in terms of direct citations and persistent influence differ the most. Finally, we introduce the diffusion process and focus on the rate of diffusion out of scientific fields, subfields and journals, and how their speeds have changed over the years.

\begin{figure}[htpb!]
\centering
\includegraphics[width =.21\columnwidth]{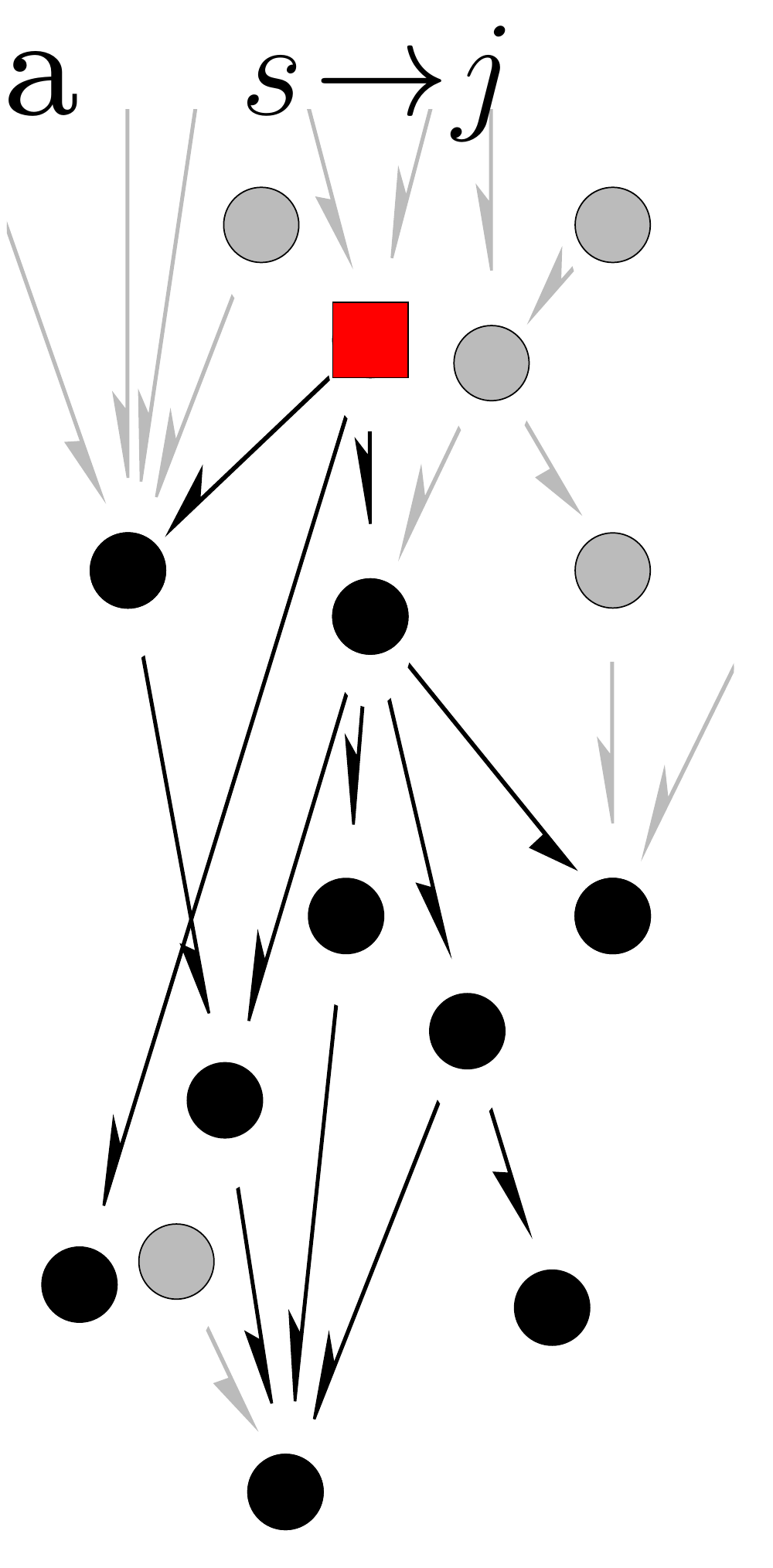}
\includegraphics[width =.21\columnwidth]{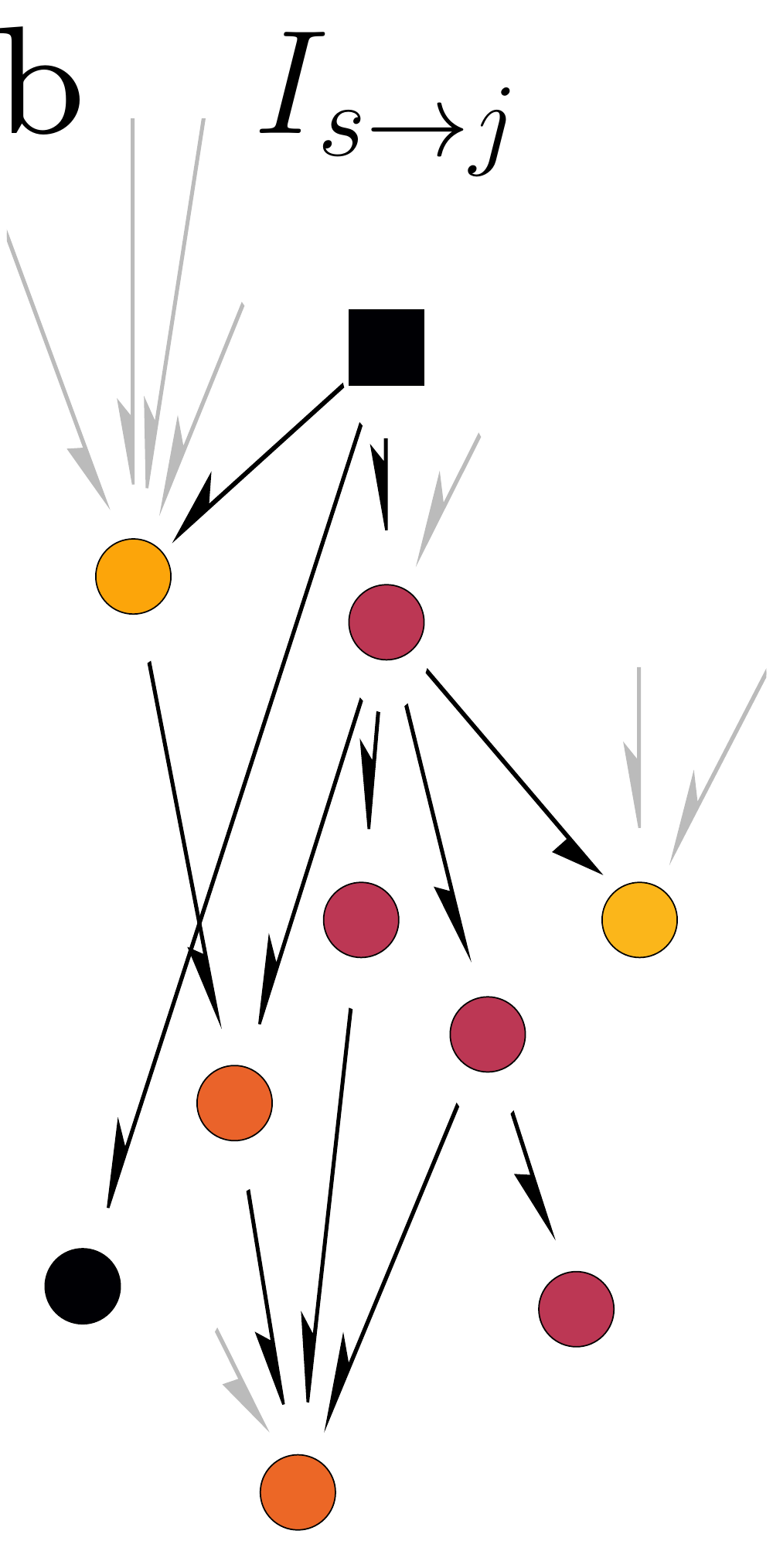}
\includegraphics[width =.21\columnwidth]{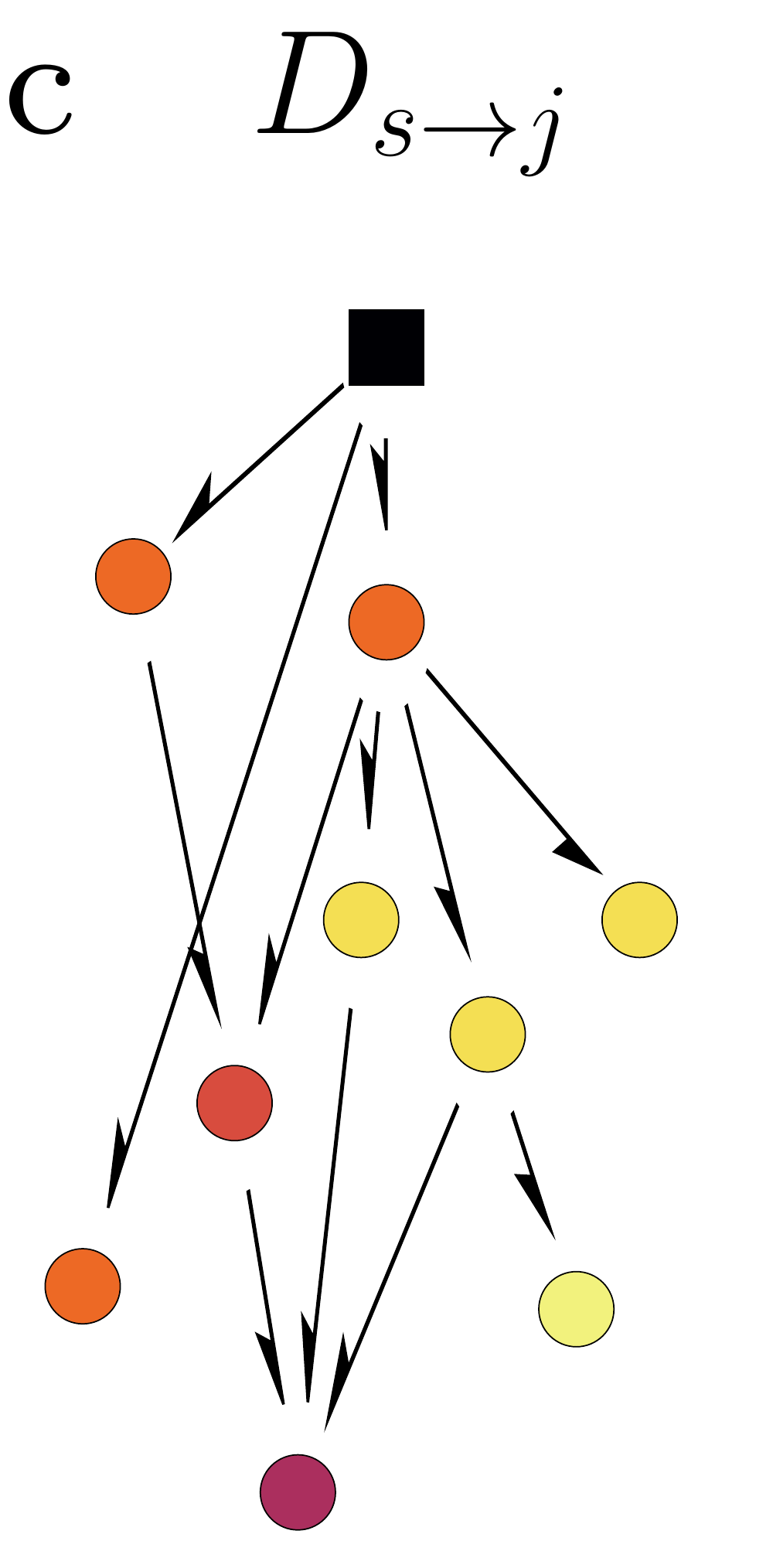}
\includegraphics[width =.21\columnwidth]{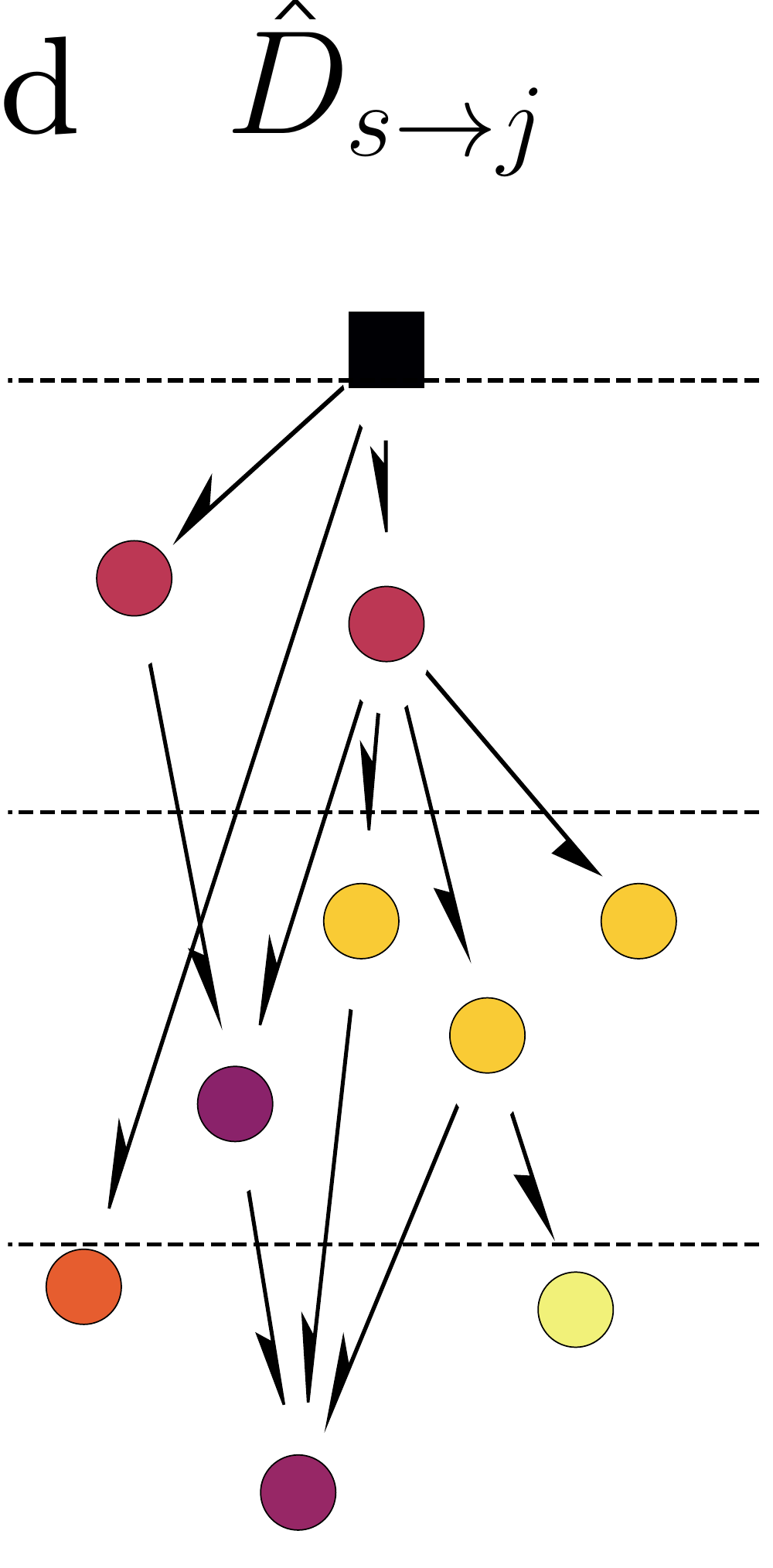}
\includegraphics[width =.10\columnwidth]{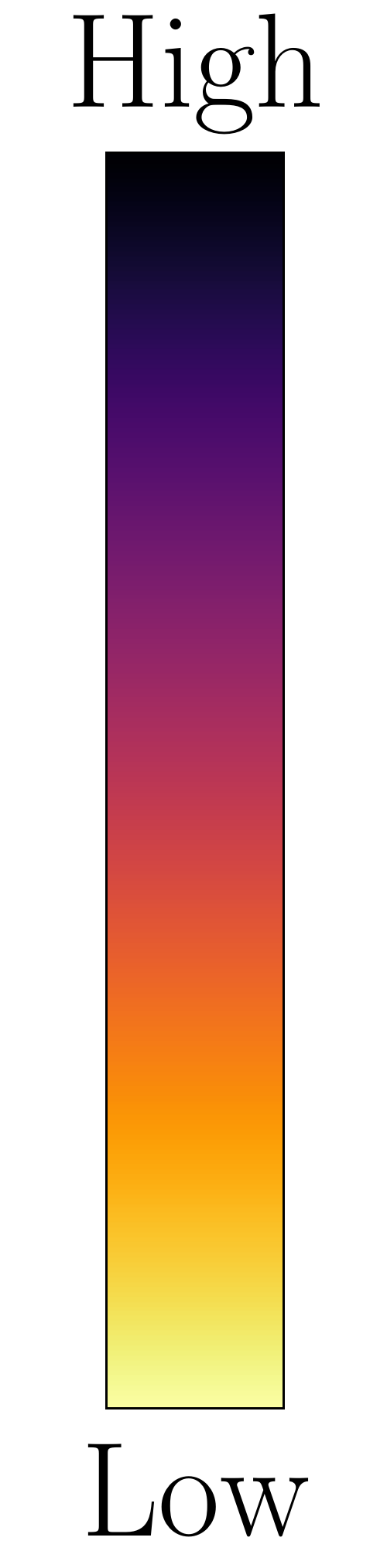}
\caption{
\footnotesize
Persistent influence and diffusion in a citation network. 
In all panels, we show the same citation network where the nodes are publications and each edge corresponds to a citation with direction corresponding to the flow of knowledge from a cited publication to a citing publication.
The square node marks the seed publication $s$.
(a) The nodes and edges \textit{reachable} from the seed node are in black and the others are in grey. 
(b) The nodes are coloured according to \textit{persistent influence} $I_{s \rightarrow j}$ (darker colours indicate higher values). Note that nodes with only a single in-edge inherit the influence value of the publication they cite, and the nodes with many citations outside of the reachable set of nodes (grey arrows) have in general low influence values. (c) The probabilities $D_{s \rightarrow j}$ for a random walker (RW) that has started from the seed node to pass each node. Note that the publications published in the future have an effect on these \textit{diffusion} probabilities: for example, the three nodes citing the seed publication each get value one third even though the last publication is published much later than the two others.
(d) The diffusion probabilities $\hat{D}_{s \rightarrow j}$ for finding the random walker in each node conditional to the random walker residing within the same time window as the node; the probabilities within each time window sum up to one.
The citation network is divided into three time windows after the seed node that are separated by the dotted horizontal lines.
Now, the publications published after the end of each time-window have no effect on the diffusion probabilities of nodes inside the time window. In the case that there are edges within a time window they are not considered when calculating the diffusion values for that window to avoid systematically giving higher values for nodes that are close to the end of the window (see the Appendix for details).
}
\label{fig:schematic}
\end{figure}

\section{Persistent influence process}
\label{sec:persistent}

Our aim in this study is to model how the scientific knowledge percolates through the network of publications citing each other. 
Since the flow of knowledge within scientific publications is difficult to measure or quantify using available data, some simplifying assumptions are required.
First, we assume that each publication is only using information that is present in the publications it cites, and the amount of intrinsic information it contains is negligible to it %
\footnote{This assumption could be relaxed by introducing a parameter describing the fraction of new information in each new paper, which would act like a damping factor. For simplicity, here we only study the case where this fraction is zero.}.
Second, we assume that each publication contains the same amount of information.
Third, we assume that each of the cited publications is equally important for the citing publication. 
Fourth, we assume that the information content of a publication can be presented as the weighted sum of information contents of the different sources, as opposed to some more complicated function. 

The above ideas are formalised in a simple persistent influence spreading process. Starting from a seed publication $s$, we attribute to it an initial value of influence $I_{s \rightarrow s}=1$, while all the other publications have an initial value of 0. We then compute the influence of the seed publication on publications published after the seed by going through the publications in chronological order. Now, each publication can be directly influenced by the seed or inherit the influence of the seed through longer chains of citations. Assuming that each of the cited publications is equally important and that the influence can be summed, the rule for calculating the persistent influence of the seed publication $s$ to publication $j$ is 
\begin{equation} \label{eq:impact}
 I_{s \rightarrow j} = \sum_{i \in \Gamma_{j}^{in}} \frac{I_{s \rightarrow i}}{ k_{j}^{in}} \,,
\end{equation}
where $\Gamma_j^{in}$ is the set of publications cited by $j$, and $k_{j}^{in}=|\Gamma_j^{in}|$ is the number of references, or the in-degree, of publication $j$. The normalisation guarantees that the sum of influence that the cited publications can have on publication $j$ is at most $1$ in the case that all the cited publications also have persistent influence values of $1$. For an illustration of this model, see Fig. \ref{fig:schematic}.

\begin{figure*}[t!]
\centering
\includegraphics[width=1.0\textwidth]{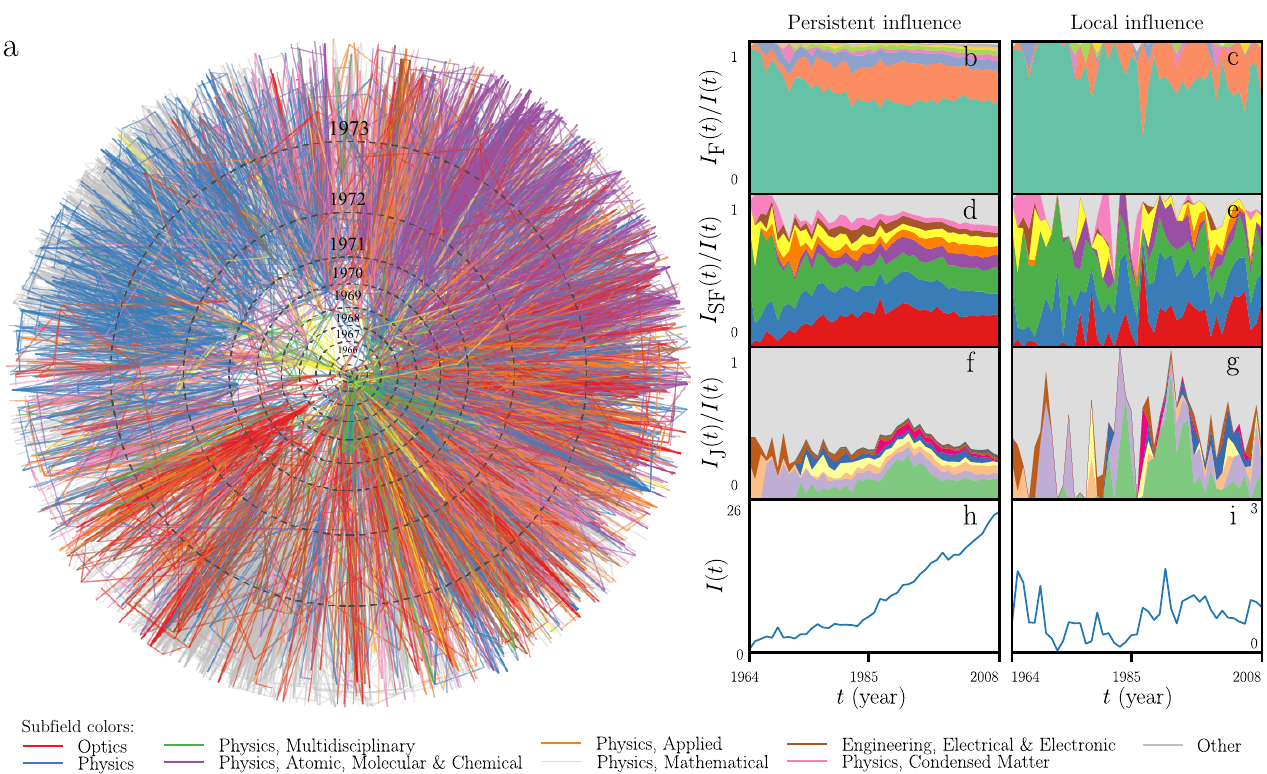}
\caption{
Persistent influence profile of Glauber's Nobel Prize winning paper from 1963 to 2008. 
Panel a: The first ten thousand, or about 9 years of, papers and citations between them in the downstream network of Glauber's paper. The edges are coloured according to the subfield of the citing paper. In case of multiple subfields, the colour is chosen according to which of these subfield has the largest area in panel d.
Glauber's paper is at the centre of the figure.
The polar coordinates of the other publications are chosen so that older publications are closer to Glauber's paper, and the angles are determined using a force-directed layout algorithm.
Panels b, d and f show the relative, aggregated persistent influence values of Glauber's paper on different groups of publications based on publication year.
Panels c, e, and g show similar yearly sums for local influence values that have been calculated using only direct citations. 
Panels b and c show the relative distribution of influence among fields (F) of science ordered in decreasing order of area in the figure starting from the bottom. The most influenced fields are \emph{physics} (green), \emph{mechanical} (orange) and \emph{engineering} (light blue)).
Similarly, panels d and e show the influence on subfields (SF) with the most influenced subfields being \emph{optics} (red), \emph{physics}, and \emph{multidisciplinary physics}, and panels f and g show the influence on journals (J) with \emph{Phys. Rev. A} (light green), \emph{Phys. Lett. A} (light purple), and \emph{Phys. Rev. Lett.} (light orange) being the most influenced ones.
Only the contribution of 8 largest fields/subfields/journals are shown and the rest is shown as light grey space.
In panels g and h, we show the total influence at the resolution of one year, \textit{i.e.}, the sum of influence values $I(t)$ for publications published in each year.
The legend on the bottom corresponds to subfields, that is, panels a, d, and e.
}
\label{fig:impact_photons}
\end{figure*}

In the presented model, each citation in the reference list of a publication is considered equally important. This feature of the model has some consequences that are important to understand.
A hypothetical publication with only one reference will draw all its influence from the cited paper as its scientific results are entirely based on that previous work. Similarly, a publication that is cited by a review publication that also cites hundreds of other publications has to share the attention with all of the other references, and only a small fraction of the information present in the cited publication is influencing the review.

When the process continues beneath the first few layers of publications in the chains of citations, typically a large number of papers become influenced by the seed node while the persistence influence values for individual papers become diluted. Now, instead of studying the persistent influence of the seed node to an individual paper, it is more meaningful to study the persistent influence of the seed publication to a group of publications.
For instance, given a seed publication, one could compute an aggregated persistent influence value $I_{\text{Physics}}(t=2019)$ by summing over the persistent influence values of all the publications published in physics journals during the year 2019.

\subsection{Persistent influence in a random citation network}
\label{sec:genmodel}

One can gain intuition on how the spreading process of persistent influence works by considering how the papers' influence scores develop in a simple model of a citation network. Let us consider a citation network where publications are published in generations, and they only cite publications of the previous generation \footnote{This assumption is for simplicity, but it is based on the reality: The number of citations a publication receives peaks few years after its publication \cite{PDBP}}.
Further, we assume that the number of publications $n(t)$ in each generation $t$ grows at a constant rate $\mu=n(t+1)/n(t)$, and the maximum values of the in- and out-degree distributions are always small compared to the current system size $n(t)$.
Now, if the in- and out degrees of papers are statistically independent, the sum of the influence values of all the papers in generation $t+1$ is on average given by:
\begin{equation} \label{eq:model}
I(t+1)=\mu p(k^{in}=0) I(t) \,,
\end{equation}
where $p(k^{in}=0)$ is the probability that a node has zero in-degree (i.e., it receives no citations). That is, the total influence of a paper to all future research grows or decays exponentially with the factor $\mu p(k^{in}=0)$ or remains constant if the generation sizes do not change and there are no ``dead-end'' publications, or if these two factors counter each other out. In reality, the scientific input has been continuously growing ($\mu > 1$) \cite{PDBP}, and thus we expect that on average the influence of early papers will grow. However, this is only the average picture. While some papers' influence values die out, others' grow faster than the average.

\subsection{Empirical results}

We will now apply the process of persistent influence to a citation network with millions of publications covering most fields of science. The data also contains information about the journal each publication is published in and the subfield classifications of the journals. This data is combined with a coarser field classification of the subfields and the information about the papers that led to Nobel Prizes \cite{PDBP}, see Appendix \ref{sec:data} for details. We begin by examining the persistent influence profile of a publication depicting how we can evaluate the cumulative influence of a paper on other papers, journals, and fields of science across time. Later, we move on to a large-scale analysis of source publications and assess the out-performance of papers related to Nobel prizes.

\subsubsection{Case study: Glauber's Nobel winning publication}

Roy J. Glauber's seminal paper on photon correlations \cite{PhysRevLett.10.84} was published in 1963, which eventually led him winning the Nobel Prize. Around nine years after the publication date, already 10 thousand publications can be connected to Glauber's paper through a chains of citations.  
The subnetwork containing these publications and the citations between them are visualised in Fig. \ref{fig:impact_photons}a. 
Different subfields in this network of persistent influence can be seen to roughly organise to their own branches, and the developments within these subfields can be approximately assessed. 
For example, publications in interdisciplinary physics journals are seen to be the early ones citing Glauber's work, but after few years the publications in the sphere of influence are more and more in the journals categorised to more specific fields of physics.

As expected, the number of publications that could be indirectly influenced by Glauber's paper in the citation network grows exponentially, reaching millions of publications before 2008 that is already a significant part of all the publications in the data set (compared to 473 direct citations). The size and topology of such a large subnetwork may be of little practical interest, but the persistent influence values are still meaningful:
Even though a massive number of publications could be influenced through chains of citations on long time ranges, most of the persistent influence values are extremely small.That is, even for the long time ranges, the influence is concentrated on a small subset of publications that can often be reached through multiple routes. By aggregating yearly influence values within the categories of publications, we can see in Fig. \ref{fig:impact_photons}b,d,f,h  that these publications are mostly on fields and subfields which are known to be impacted by Glauber's work.

To illustrate the difference between the local viewpoint of counting direct citations and the persistent influence, we show a local influence profile calculated using Eq. \ref{eq:impact} but only considering direct citations to the Glauber's original publication in Figs. \ref{fig:impact_photons}c,e,g,i.
At first sight, the most notable difference between the global and local profiles is that the global profile is much smoother than the local one. This is expected as if a field was influenced by a seed article, the articles from the field are also likely to cite the seed. However, a closer inspection reveals more subtle differences between the global and local profiles. The influence of Glauber's paper on publications categorised to fields (Fig.\ref{fig:impact_photons}b) shows a strong persistence in \emph{physics} with a gradually growing contribution to fields called \emph{mechanical} and \emph{engineering}. This pattern is already stable after ten years in the persistent influence, but it is not visible in the local profile even though the second most influenced field, \emph{mechanical}, starts to cite the seed publication later on.
A similar effect of \emph{optics} becoming prominent in the persistent influence much before it is picked up by the local influence profile 
 can be seen when looking at the categorisation in the level of subfields (Fig. \ref{fig:impact_photons}d-e).
Going down to the level of journals we can see that the contribution to \emph{optics} is mainly due to two journals: \emph{Physical Review A} and \emph{Optics Communications}, which once again are not as strongly present in the local profile (Fig. \ref{fig:impact_photons}f-g). 

The total persistent influence of Glauber’s publication on publications published during each year grows (except for few years), but the local influence remains relatively stationary when one considers only direct citations. This observation is qualitatively in line with what would be expected based on the generational model defined in Section \ref{sec:genmodel}, because the persistent influence values are sensitive to the expansion in science that has happened at a rapid pace since 1963. However, it is not clear from this analysis if the rate at which the persistent influence grows is a typical example of a publication of that time or if this Nobel winning publication is somehow special.

\begin{figure}[t!]
\centering
\includegraphics[width=.99\columnwidth]{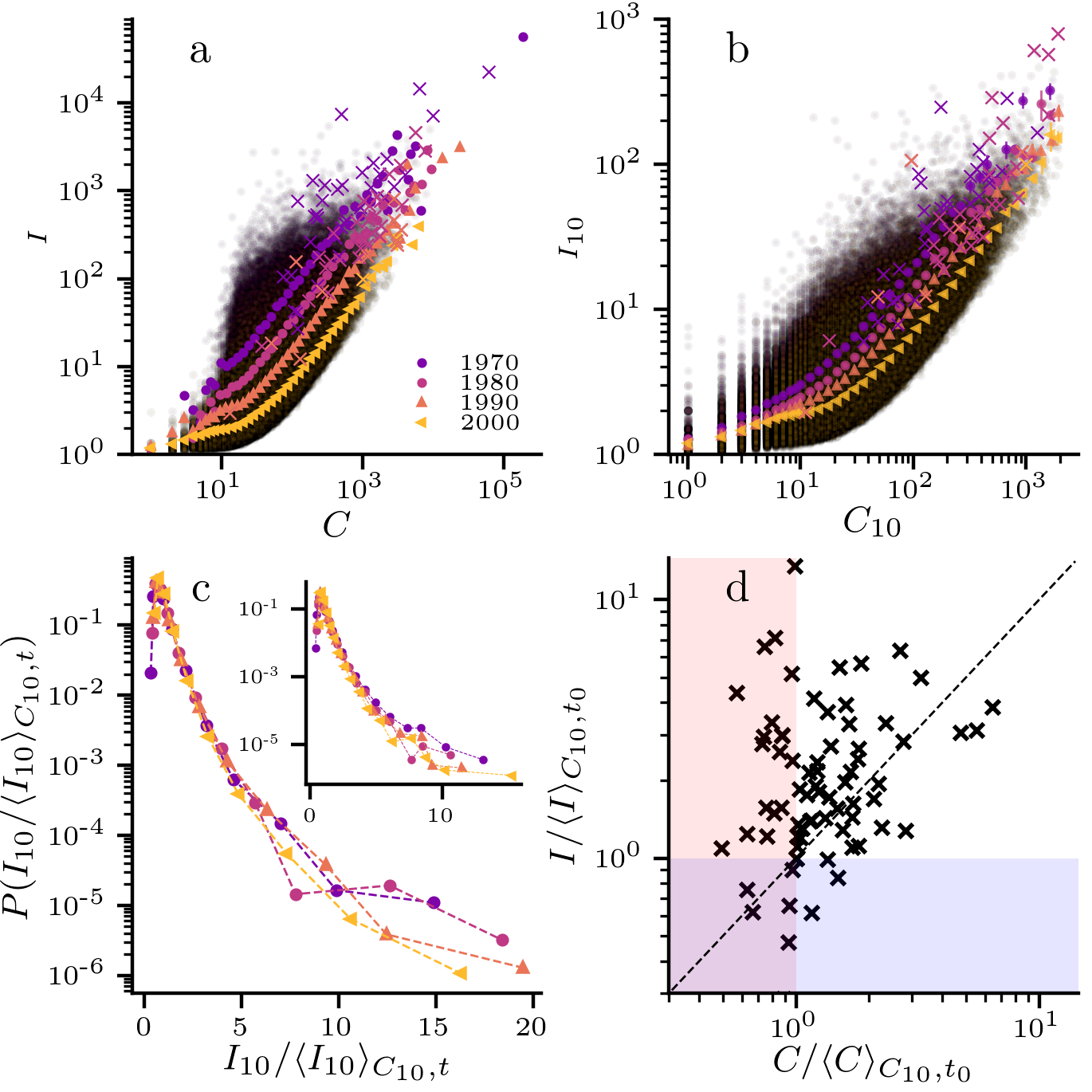}
\caption{
Relationship between citation counts $C$ and persistent influence $I$, and the outperformance of Nobel papers.
(a) The total persistent influence of a paper on average increases as a function its citation count, but citation counts alone can not explain the persistent influence values. Black dots correspond to individual papers, and the coloured markers show the total persistent influence value averaged over papers with similar citation counts and same publication year, $\langle I \rangle_{C,t}$. 
The crosses represent Nobel papers with the colours chosen according to the closest 10 years. 
(b) The same as in panel a, but with citations and persistence influence being aggregated only for the first ten years after the publication of each paper: $C_{10}=C(\Delta t < 10y)$ and $I_{10}=I(\Delta t < 10y)$.
The average curves thus correspond to $\langle I_{10} \rangle_{C_{10},t}$.
(c) The distribution of total persistent influence values divided by the average total persistent influence of a group of papers with similar order of magnitude of citations ($\approx100$). 
The inset shows the same distribution for publications with one order of magnitude less citations.
The colour coding is the same as in panels (b) and (c).
These corresponds to the distribution of aggregated persistent influence values for vertical slices of panel b.
(d) The outperformance of Nobel-winning papers in total number of citations $C / \langle C \rangle_{C_{10}, t_0}$ and total persistent influence $I/ \langle I \rangle_{C_{10},t_0}$ when compared to a group of reference papers with similar publication years ($t_0$) and numbers of citations in the first 10 years ($C_{10}$).
Now Nobel papers outperform their peers in 70\% of the cases in terms of citations, but outperform their peers at even a higher rate (87\%) in terms of persistent influence.
}

\label{fig:impact_panel_full}
\end{figure}

\subsubsection{System level analysis of seed publications} %

In order to gain insight into the relationship between the persistent influence and local citation count at the system level, we repeated the influence profile calculation described above for all the papers in our dataset published between 1970 and 2008 and having at least 20 citations. We chose 20 as the required number of citations primarily to limit the computational costs of our analyses while focusing on papers with significant scientific impact. In total, this amounts to approximately $6.2$ million seed publications. To summarize the results, we focus on the total persistent influence that each publication gathers ($I=\sum_t I(t)$).

There is a strong positive correlation between the total number of citations $C$ and the total persistent influence $I$ a publication receives (Fig.\ref{fig:impact_panel_full}a). However, at the same time there is large variation, up to several magnitudes, in the total persistent influence values of publications that receive similar numbers of citations. This dependence between the local citation counts and the globally computed persistent influence values resembles the results in Ref. \cite{Chen20078} obtained for another global measure of publication importance. Further, this result indicates that the number of citations \textit{per se} is not sufficient to fully summarize the persistent influence that a single paper has had within the scientific literature.
Also, we see that older papers manage to gather a significantly higher amount of impact with the same number of citations. This is expected, as the older papers have had more time to gather cumulative influence among their scientific off-springs, and may also benefit from the growing system size.

To remove this advantage of the older papers, we calculate the total persistent influence that each paper gathers during the first 10 years since its publication date $t_0$ ($I_{10}=\sum_{t-t_0 \leq 10 \text{\,years}} I(t)$). We have chosen to use 10 years as the limit so that there would be sufficient time for publications to gather citations and persistent influence. Note however, that if 5 years were used as the limit instead, the results would remain qualitatively similar.
This partly removes the effect of the publication year, but older publications with the same number of citations are still seen to have slightly higher total persistent influence as compared to the newer ones (Fig.\ref{fig:impact_panel_full}b).
The remaining difference between the publication years might be due to other changes in the citation practices such as the increase in the length of reference lists, which causes the denominator in Eq.\ref{eq:impact} to reduce the amount of influence to a single citing paper.
However, the distributions of persistent influence values of publications with the same number of citations within 10 years ($C_{10}$) have super-exponential tails that are relatively independent of the publication year (Fig.\ref{fig:impact_panel_full}c).
This shows that even on a time scale of 10 years, publications having similar numbers of citations can have varying scientific impacts when measured using the total persistent influence.

\subsubsection{Performance of Nobel-winning publications}

What are the publications that achieve higher persistent impact values than expected based on their citation count and are these publications somehow more important for science than the others? To begin to answer these questions, we studied 74 papers that were associated with a Nobel Prize \cite{NobelAge}, which we assume here is a hallmark of scientific importance.

The Nobel prize winning papers mainly populate the top-right corner in Fig.\ref{fig:impact_panel_full}a-b, indicating that, generally, they tend to gather high absolute values of citations and influence. However, as many of them are among the few most cited papers in the data their persistent influence is not significantly higher than other papers with similar citation counts.
Because the Nobel prizes are often given with a significant and growing delay \cite{NobelAge} indicating that their significance is acknowledged only long after the original research is published, 
we divert our attention to the longitudinal aspect of the influence process. The idea behind this approach is to take a set of control papers which, at a certain point in time, are apparently equivalent to the Nobel paper and to compare the performance of the Nobel paper with respect to the control group at a later point in time. Practically speaking, we construct a set of control papers that have been published in the same year, $t_0$, as the reference paper and have at most $10\%$ difference in the citation count after 10 years, $C_{10}$. 
We calculate the out-performance of a Nobel paper to its controls as the ratio $I/\langle I \rangle_{C_{10},t_0}$, where $I$ is the total persistent influence value of the Nobel winning publication and $\langle I \rangle_{C_{10},t_0}$ is the average total persistent influence of the control papers. These out-performance values can be compared to the outperformance of Nobel papers in terms of total number of citations  $C / \langle C \rangle_{C_{10},t_0}$: Out of 74 Nobel papers 65 outperform their controls in persistent influence and 50 of them outperform the controls in citations (see Fig. \ref{fig:impact_panel_full}d), which makes the persistent influence significantly more likely to display outperformance than citation counts (p-value $<10^{-4}$ for the hypothesis that 50/74 chances could have produced 65 or more successes). Furthermore, in 73$\%$ of the cases (54 papers out of 74) the persistent influence outperformance is greater than the citation one (again, the p-value $<10^{-4}$ for the hypothesis that this difference was produced with a null model that gives equal probability for the persistent influence and citation count to outperform the other). 

Based on the above numbers and the overall picture presented in Fig. \ref{fig:impact_panel_full}d, it can be said that the Nobel papers outperform the controls in terms of persistent influence and citation counts, and that typically the outperformance of the Nobel papers is greater in terms of the persistent influence than citation counts. For some Nobel papers, the outperformance is smaller in terms of persistence influence but the difference to citation-count-based outperformance is then typically small. An interpretation of these results is that the Nobel papers often constitute the starting points of new growing areas of science that reach beyond the publications directly citing them, as one would expect from publications with influential results.
This conclusion supports the finding of \cite{10.1371/journal.pone.0113184}, where the groundbreaking papers by important Nobel laureates were found to have large networks of child nodes spanning over multiple layers.

\subsubsection{Using rankings to detect influential but low-cited papers}

\begin{figure}[t!]
\centering
\includegraphics[width =.99\columnwidth]{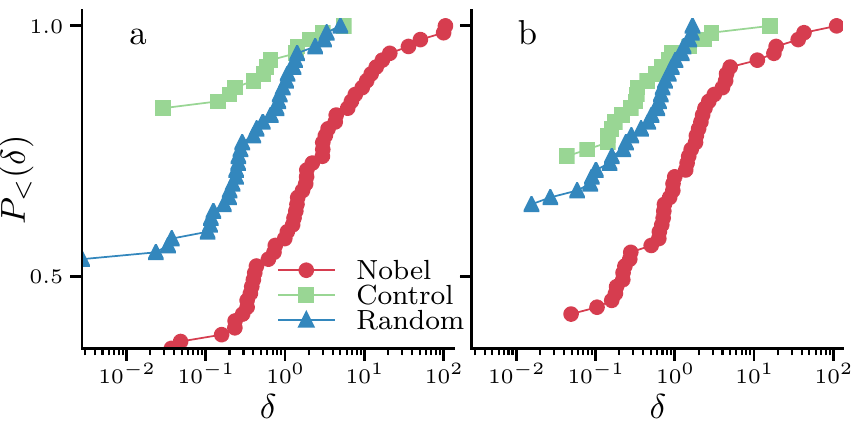}
\caption{
Nobel papers have higher ranks in terms of persistent influence than citations.
(a) Cumulative distributions of relative difference in persistent influence rank and citation count rank, $\delta = \frac{R_C - R_I}{R_I}$.
Only the positive values of $\delta$ are shown, and the amount of probability mass in the negative side of the distribution can be read from the smallest shown value. 
Three categories of papers are considered: Nobel papers (red circles), Nobel control set of papers having the same publication year and being within $3\%$ in citation volume (green squares), and papers selected uniformly at random from all papers (blue triangles).
In case no papers were found within the 3\% interval, the most similar paper in terms of citation was used as the control.
(b) The same distributions as in (a) but using persistent influence values and citation counts after 10 years (i.e., $I_{10}$ and $C_{10}$ instead of $I$ and $C$).
}
\label{fig:nobel_ranking}
\end{figure}

The raw values of the persistent influence and citation counts have fat tails, which can make the analysis very sensitive to outliers, and also depend heavily on the publication year.
Instead of looking at these raw values, we will now use ranks as a more robust measure of publication importance. To this end, we order the papers published within the same year according to the citation count and persistent influence, and give each paper ranks $R_C$ and $R_I$ in terms of the citations and persistent influence, respectively, such that small rank means high value.
We analyse the ranks by the publication year to avoid the biases due to higher values of citation counts and persistent influence for older papers.
Now, we are primarily interested in identifying those publications that have high persistent influence values but have received relatively few citations over the years.
To this end, we define the relative change in ranking as
\begin{equation}
 \delta = \frac{R_C - R_I}{R_I}\,.
 \label{eq:deltarank}
\end{equation}
Here we divide the absolute difference in rankings by $R_I$ to emphasize the publications that have low ranks in terms of the persistent influence: given two publications with the same absolute difference in ranks, the higher $\delta$ is given to the publication with more persistent influence.

Fig.\ref{fig:nobel_ranking}a shows the cumulative distribution of $\delta$ values at different times for the 74 Nobel papers, a control group of papers within $3\%$ of citations of each Nobel paper and uniformly randomly selected papers. The Nobel publications are seen to be much more likely to be among the publications that have high persistent influence but relatively low citation count than publications in either of the control groups. This effect persists when we use the 10-year persistent influence values $I_{10}$ and citation counts $C_{10}$ for computing the ranks (Fig.\ref{fig:nobel_ranking}b).

The Nobel publications are overrepresented in the publications with high $\delta$ values, but what are the publications with the largest relative difference in the persistent influence and citation counts? Table \ref{tab:dead_beauties} shows the publications with largest $\delta$ values for all the papers in our dataset. Some of these publications might be ``forgotten beauties'' in the sense that they have contributed to the development of several other important publications but were forgotten as these more prominent publications gathered all the citations related to the breakthroughs that were made.
Indeed, such forgotten publications appear in the list.
The publication with the highest $\delta$ score (\#1) is related to DNA hybridisation techniques and is cited by several early publications in the emerging field of molecular biology, most notably by E.M. Southern's work on the \textit{Southern Blot}, a widely-used method for identifying specific DNA sequences from DNA samples.
The second, third and fifth are in the reference list of Sanger's Nobel Paper \textit{DNA sequencing with the chain terminating inhibitors} which had a massive impact on the biological and medical sciences. 
Articles \#5, \#9, \#15, and \#21 are all linked to the identification, classification, or prediction of very well known diseases (Prostate Cancer, AIDS, Leukemia).

As expected, Biology is the main contributor to the list, due to it being the largest field in our dataset. However, unlike for the highest cited papers for each year where Biology is virtually the only field present (see Table \ref{tab:highest_citations} in the Appendix), in this list diverse aspects of science are included. There are many papers from physics, with \#4, \#12, and \#24 being linked to the discovery of High Temperature Superconductivity, while \#11 and \#30 are linked to the development of Carbon Nanotubes and, in general, of Material Science. \#10 is among Amano's works that led to his Nobel Prize for the invention of efficient blue light emitting diodes. \#25 is a small summary of the recent (at the time) discoveries in the mathematical field of Fractals, as it was among the few cited works in the famous \textit{Self Organized Criticality: An Explanation of 1/f Noise}. Also, we can see contributions from Economics and Engineering with \#27, which discusses a computer method able to improve the efficiency of production of industrially assembled products. \#28 is one of the earliest attempts of statistical methods for assessing agreement between different clinical measurements. 

Finally, the list also shows evidence of the relatively new field of complex networks, with \#23 being among the earliest papers in the field and cited by virtually all the most significant early publications in the field. Overall, we can see how $\delta$ is able to describe the growth in the whole scientific field of certain discoveries/subfields/hot topics by being able to identify
low cited papers that have been crucial in their early stages.

\begin{table}[t!]\footnotesize
\caption{The 30 publications with the highest relative difference in influence rank and citation count rank $\delta$ given in Eq.~\ref{eq:deltarank}.}
\label{tab:dead_beauties}

    \begin{tabular}{c c c|p{6.5cm}}    
\# & $R_c$ & $R_I$ & Title (year) \\ \hline
1 & 37588 & 5 & {\tiny \it Hybridization On Filters With Competitor Dna In Liquid-Phase In A Standardand A Micro-Assay (1974) }\\  
2 & 23366 & 5 & {\tiny \it Nucleotide And Amino-Acid Sequences Of Gene-G Of Phix174 (1976)}\\  
3 & 62269 & 18 & {\tiny \it Invitro Polyoma Dna-Synthesis - Inhibition By 1-Beta-D-Arabinofuranosyl Ctp (1975)}\\  
4 & 88381 & 32 & {\tiny \it Inhomogeneous Superconducting Transitions In Granular A1 (1980)}\\  
5 & 26353 & 10 & {\tiny \it  An Adjustment To The 1997 Estimate For New Prostate Cancer Cases (1997)}\\  
6 &  28047 & 11 & {\tiny \it Molecular Hybridization Between Rat Liver Deoxyribonucleic Acid And Complementary Ribonucleic Acid (1970)}\\  
7 & 63260 & 25 & {\tiny \it A Novel Method For The Detection Of Polymorphic Restriction Sites By Cleavage Of Oligonucleotide Probes - Application To Sickle-Cell-Anemia (1985)}\\  
8 & 105590 & 42 & {\tiny \it Phase-Diagram Of The (Laalo3)1-X (Srtio3)X Solid-Solution System, For X-Less-Than-Or-Equal-To 0.8 (1983)}\\  
9 & 131750 & 72 & {\tiny \it A New Method Of Predicting Us And State-Level Cancer Mortality Counts For The Current Calendar Year (2004)}\\
10 & 114723 & 67 & {\tiny \it  Zn Related Electroluminescent Properties In Movpe Grown Gan (1988)}\\  
11 &  26020 & 16 & {\tiny \it Structure And Intercalation Of Thin Benzene Derived Carbon-Fibers (1989) }\\  
12 &  12231 & 8 & {\tiny \it The Oxygen Defect Perovskite Bala4Cu5O13.4, A Metallic Conductor (1985)}\\  
13 &  19801 & 13 & {\tiny \it Translation Of Encephalomyocarditis Viral-Rna In Oocytes Of Xenopus-Laevis (1972)}\\
14 &  4216.5 & 3 & {\tiny \it Amplified Ribosomal Dna From Xenopus-Laevis Has Heterogeneous Spacer Lengths (1974)}\\  
15 &  42143 & 30 & {\tiny \it  Classification Of Acute Leukemias (1975)}\\  
16 &  62485 & 48 & {\tiny \it Wild Topology, Hyperbolic Geometry And Fusion Algebra Of High Energy Particle Physics (2002)}\\  
17 & 58242 & 46 & {\tiny \it Relation Between Mobility Edge Problem And An Isotropic Xy Model (1978)}\\  
18 &  42981 & 34 & {\tiny \it Transcriptional And Posttranscriptional Roles Of Glucocorticoid In The Expression Of The Rat 25,000 Molecular-Weight Casein Gene (1986)}\\  
19 &  114240 & 91 & {\tiny \it The Use Of Biotinylated Dna Probes For Detecting Single Copy Human Restriction-Fragment-Length-Polymorphisms Separated By Electrophoresis (1986)}\\  
20 &  92031 & 74 &  {\tiny \it A Solid-State Nmr-Study On Crystalline Forms Of Nylon-6 (1989)}\\  
21 &  89271 & 74 & {\tiny \it Multiple Opportunistic Infection In A Male-Homosexual In France (1982)}\\  
22 &  11535 & 10 & {\tiny \it Synthesis Of Ribosomal Rna In Different Organisms - Structure And Evolution Of Rrna Precursor (1970)}\\ 
23 & 11227 & 10 & {\tiny \it Small-World Networks: Evidence For A Crossover Picture (1999)}\\  
24 & 12064 & 13 & {\tiny \it Superconductivity At 52.5-K In The Lanthanum-Barium-Copper-Oxide System (1987)}\\  
25 &  71919 & 82 &  {\tiny \it Fractals - Wheres The Physics (1986)}\\  
26 &  28044 & 33 & {\tiny \it The Complete Structure Of The Rat Thyroglobulin Gene (1986)}\\  
27 &  21222 & 25 &{\tiny \it Interference Detection Among Solids And Surfaces (1979)}\\  
28 &  81374 & 99 & {\tiny \it Comparison Of The New Miniature Wright Peak Flow Meter With The Standard Wright Peak Flow Meter (1979)}\\  
29 & 7962 & 10 & {\tiny \it Studies On Polynucleotides .105. Total Synthesis Of Structural Gene For Analanine Transfer Ribonucleic-Acid From Yeast - Chemical Synthesis Of An Icosadeoxyribonucleotide Corresponding To Nucleotide Sequence 31 To 50 (1972)}\\  
30 &  26908 & 34 & {\tiny \it Materials Science - Strength In Disunity (1992)}\\ \hline 
       \end{tabular}
\end{table}

\section{Diffusion}

The persistent influence spreading method we just introduced is a simple and elegant method to model the spreading of knowledge in the citation network, but it is not the only plausible one. In the influence spreading, the tracked quantity can be copied, and the total amount in all publications can grow.
Next, we instead define a diffusion method where the original mass placed on the seed node (or nodes) is always strictly conserved. This allows us to track the diffusion of ideas not only from single publications but across journals, subfields, and fields.

The idea behind the diffusion method is to start a random walker from a seed publication that is randomly selected from a set of seed publications, and then, at each step, let it move from a publication to future publications citing it (see Fig. \ref{fig:schematic}). That is, if we have $N$ seed papers we assign the same initial probability $\hat{D}_{s \rightarrow s} = 1/N$ to all of them, and calculate the probability that the random walker goes through the rest of the publications as
\begin{equation} \label{eq:diffusion}
 \hat{D}_{s \rightarrow j} = \sum_{i \in \Gamma_{j}^{in}} \frac{\hat{D}_{s \rightarrow i}}{ k_{i}^{out}} \,,
\end{equation}
where $\Gamma_j^{in}$ is the set of publications cited by $j$, and $k_{i}^{out}$ is the number of publications citing $i$ (or, the out-degree).

Note that this process is sensitive to the time window we choose, as the future publications that we do not know about yet will change the degrees $k_{i}^{out}$, and thus probabilities of trajectories of the walkers. To negate this effect, we will focus on walks that have not passed beyond our observation year. That is, we only use the information available of each observation year, and the random walk process is re-calculated for each starting year and observation year pair. In addition, we are not interested in the raw probabilities of random walkers visiting the nodes $\hat{D}_{s \rightarrow i}$, but on the conditional visiting probabilities $D_{s \rightarrow i}(t)$ given that the walker is within given year $t$. Further, we will disregard any citations within each observation year to avoid emphasising the publications that are published towards the end of each year. It is also possible to define the model such that both the distribution of the initial mass and the diffusion process are based on the citations of the seed papers and of the child nodes. The results, although qualitatively different, are quantitively identical. See Appendix \ref{sec:alt_diffusion} for details about these methods and Appendix \ref{sec:computations} for details of how the diffusion probability is calculated.

\begin{figure}[htpb!]
\centering
\includegraphics[width=.99\columnwidth]{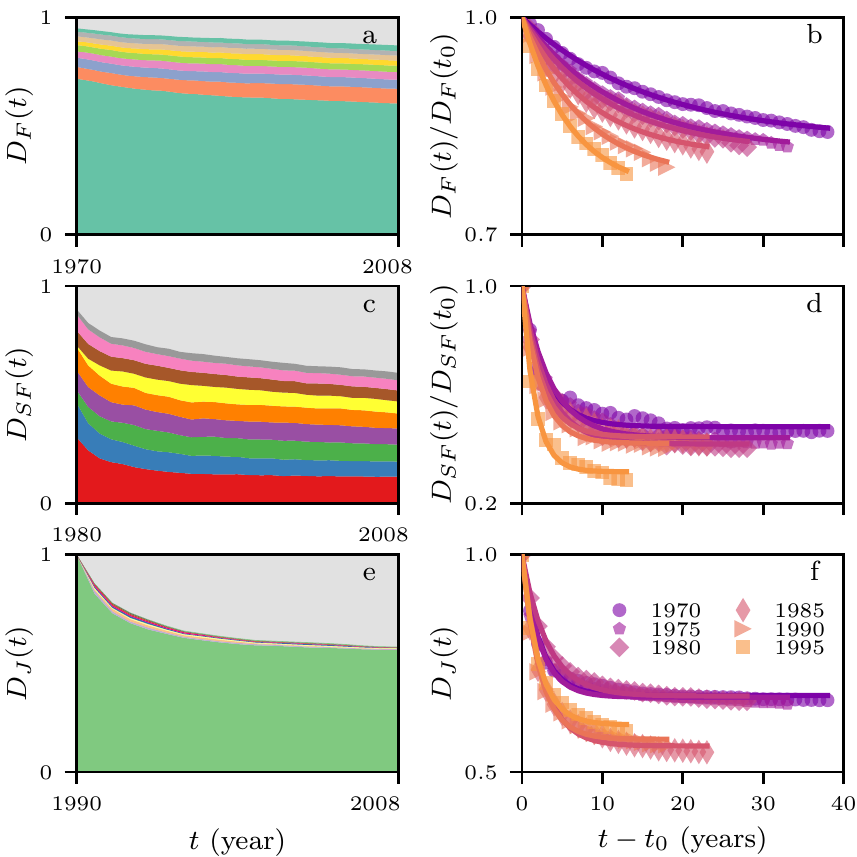}
\caption{
Examples of the diffusion process (a,c,e) and the exponential decays to the metastable state (b,d,f). 
(a,b) The diffusion of scientific value for (a, b) \emph{economics} in 1970, (c, d) \emph{evolutionary biology} in 1980, and (e, f) the \emph{British Medical Journal} in 1990. 
(a,c,e) The area in the bottom shows the amount of scientific value retained by the initialisation field (F), subfield (SF) or journal (J).
The contribution of the 8 largest fields/subfields/journals are also shown individually, and the contribution of other fields/subfields/journals are shown as light grey space.
In panel c the colours represent the following subfields (from bottom to top): \emph{evolutionary biology}, \emph{biology}, \emph{miscellaneous}, \emph{plant sciences}, \emph{anthropology}, \emph{ecology}, \emph{zoology}, \emph{genetics \& heredity}, \emph{arts \& humanities}, and \emph{general biology}.
In panel e the colours represent the following journals (from bottom to top): \emph{Br. Med. J.}, \emph{Lancet}, \emph{Br. J. Gen. Pract.}, \emph{Bmj-British Medical Journal}, \emph{Med. J. Aust.}, \emph{Postgrad. Med. J.}, \emph{Arch. Dis. Child.}, \emph{J. Clin. Pathol.}, and \emph{Med. Clin.}.
The panels (b,d,f) instead show the renormalised value of $D$ retained within each field/subfield/journal for different years (markers) and with the exponential fitting (solid line).}
\label{fig:push_panel}
\end{figure}

\subsubsection{Relationship to persistent influence}

The equation for diffusion, Eq. \ref{eq:diffusion}, resembles the equation for the persistent influence, Eq. \ref{eq:impact}, but the similarity between these two runs deeper than that. In fact, the persistent influence values $I_{s \rightarrow i}$ can also be considered in terms of a diffusion process that goes backwards in time. That is, the persistent influence value starting from a single seed publication $s$, $I_{s \rightarrow i}$, is equal to $\hat{D}_{i \rightarrow s}$ in a network where the directions of the links are reversed and the diffusion process starts from the node $i$. In this sense, the two processes, the persistent influence and the diffusion are complementary to each other.

\subsection{Empirical Results}

When starting the diffusion process, we select a field, a subfield, or a journal and a starting time $t_0$, and track the probability that the random walker is found in a field $D_{F}(t)$, a subfield $D_{SF}(t)$, or a journal $D_{J}(t)$ at time $t$ (using a resolution of one year). This process is illustrated in Fig. \ref{fig:push_panel}a,c,e for the initial field of \emph{economics} (in 1970), the subfield of \emph{evolutionary biology} (in 1980), and the \emph{British Medical Journal} (in 1990). In the case of journal-level aggregation, the probability for finding the random walker in the seed journal equals one in the beginning. However, when aggregating results on the level of fields or subfields, papers can be associated with multiple fields or subfields. In this case, we initially split the total probability equally among all the fields or subfields the paper belongs to.

As the diffusion process progresses, the random walker jumps between publications that can be in different groups (i.e., fields, subfields, or journals) with the possibility of returning to the original one. 
However, as expected in any diffusion process, the random walker will forget its origin, and the amount of probability mass in the original group goes down monotonically as time progresses. This process is observed to be slow in the examples of Fig. \ref{fig:push_panel}a-c. 
For \emph{economics}, the probability of finding the walker in other fields has not grown significantly even after 40 years of random walk. 
Similarly, for the subfield of \emph{evolutionary biology} and \emph{British Medical Journal} the walker is far from forgetting its origin during the observation period. These two initial groups seem to lose probability mass very fast to other groups in the beginning, but this fast phase is then followed by a very slow change, which is almost like a plateau compared to the initial rate of change.

In a stationary system, which the citation network is not, one would expect the diffusion eventually to completely forget its origin and settle to a stationary distribution. Furthermore, such stationary citation system would likely be ergodic, and the stationary state would be unique. The behaviour observed for the fields, subfields and journals do not reach such a unique stationary state, but they seem to be rapidly reaching a metastable state, where the probability of finding a random walker in a specific field, subfield or journal does not change much in time but where these probabilities are not independent of the origin. This observation could, for example, be explained by a network structure containing strong clusters where the random walkers get trapped for the time scales of the data.

\subsection{Summarizing the diffusion curves}

The above procedure of observing the diffusion patterns is very cumbersome if the goal is to get an idea of the system-wide behaviour of diffusion starting from different groups. In order to summarize the results of the patterns such as the ones presented in Fig. \ref{fig:push_panel}a,c,e, we have looked only at the amount of value retained by each group. For each starting year $t_0$, we observe the yearly values of the initial group $D_{G}(t)$ and normalize them with the initial value before any diffusion $D_G(t_0)$. We then fit each curve with an exponential of the form: 
\begin{equation} \label{eq:curve}
D_G(t)/D_G(t_0) = (1-\beta) e^{-\alpha t} + \beta\,, 
\end{equation}
which follows well the typical shape we observe for the curve. The fits for the previously discussed example cases are shown in Fig. \ref{fig:push_panel}b,d,f.
These fits allow us to summarize both the rate of change of the value of the initial years (through $\alpha$) and find the final plateau value (through $\beta$). Therefore, $\alpha$ can be used to measure the rate at which one group shares its knowledge with other groups that are close to it, and $\beta$ instead represents the intrinsic ``conservativeness'' of a group and the other groups related to it, i.e. the amount of knowledge retained within the boundaries of the group itself in medium time scales. 

In order to provide an easier metric for the exponential decay, we introduce a related parameter called half-life $t^{1/2}$ defined as the time required to lose half of the possible plateau value. That is by solving %
\begin{equation} \label{eq1}
\begin{split}
(1 - \beta)e^{-\alpha t^{1/2}} + \beta & = 1- \frac{1 - \beta}{2} \,,
\end{split}
\end{equation}
we get the conventional definition of half-life:
\begin{equation} \label{eq:half_life}
t^{1/2} = \ln(2)/\alpha \,.
\end{equation}

\subsection{$t^{1/2}$ and $\beta$ in the data}

With the above-mentioned ideas on summarizing the diffusion processes in mind, we can put together information about all the possible initial times and  fields, subfields, and journals. Table \ref{tab:fields_1970} shows the values for the half-lives, $t_{1/2}$s, and plateau values, $\beta$s, in 1970 and 1995. We can see that in general there is a decreasing trend for half-lives while the plateau value $\beta$ instead shows a much more stable pattern. 

Some of the individual fields display interesting patterns. The field of \emph{multidisciplinary} has the lowest half-life for both starting years, coherently with the fact that it is meant to be a field open to sharing its knowledge with others. However, its change in $\beta$ is positive and the second highest (behind \emph{music}). 

This could indicate that \emph{multidisciplinary} might have become a field of its own, which can retain random walkers within itself for long periods of time. This observation is coherent with the evidence that shows the increasing role of interdisciplinarity in science \cite{Pan2012interdisciplinarity,Sinatra2015,Rosvall29012008,Porter2009}. It is also interesting to note that while in 1970 some humanistic fields show very high values for their half-lives, e.g. (\emph{philosophy}, \emph{history}, \emph{anthropology}, \emph{literature}, and \emph{linguistics}), these fields also show some of the highest changes in time, putting them much closer to hard sciences these days than they were before.

A more systematic observation on the changes in the speed of the diffusion process is shown in Fig. \ref{fig:distribution_panel}(a,c,e) which displays the change of half-lives for a set of fields and changes in the distributions of half-lives for all subfields and journals. All of the fields show a speeding-up pattern, losing between 20 and 60 percent of their half-life values, while for subfields and journals the more recent cumulative distributions of half-lives are above the older ones, showing that the values have in general decreased.

\subsection{Renormalizing the time}

Previous studies show \cite{PDBP} that the time as such may not be the best choice for measuring the rate at which changes happen in science, but instead  use the numbers of papers published as the measure of progress. In other words, science can be considered to be ``updated'' every time a new publication is introduced in the system, and the count of such updates is a better measure of ``progress'' in science than simply the time elapsed.

As the number of publications per year $N(t)$ grows exponentially, one would theoretically expect that the functional form of the exponential decay given by~\ref{eq:curve} would have to be adjusted when measuring time in terms of published papers. However, as the growth rate of yearly publication number is sufficiently small $N(t) \approx
N_{0}e^{\delta t}$ with $\delta \sim 0.05 \, (\text{year}^{-1})$ across all the fields, we can approximate this exponential growth by a simple, linear function $e^{\delta t} \approx (1+\delta t)$ for the span of time $t$ that we consider.
This allows us to use the functional forms given by Eq.~\ref{eq:curve} also for fitting the half-lives $t^{\frac{1}{2}}$ and the asymptotic fractions $\beta$ when the time is measured in terms of published papers.

Hence we are able to follow the earlier procedure we applied to real time and quantify the half-lives in terms of the numbers of published papers. We chose to use the number of published papers in each field as a unit of time for the fields, while for the subfields and journals we used the data from all scientific publications.

\begin{figure}[htpb!]
\centering
\includegraphics[width=.99\columnwidth]{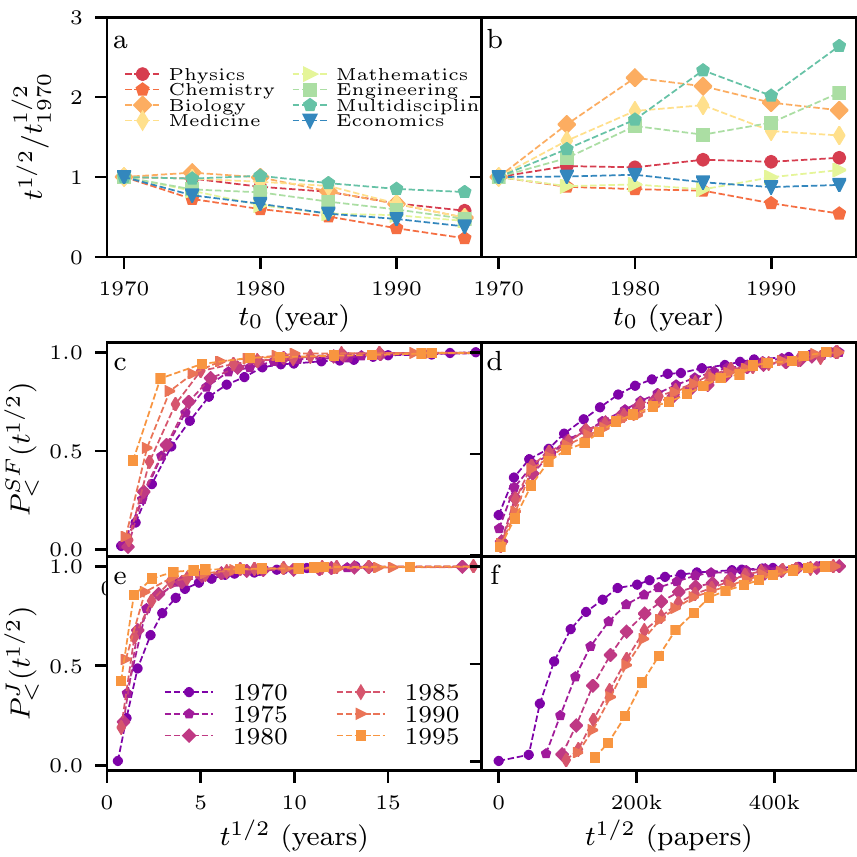}
\caption{Changes in half-lives in real time (left column, panels a,c,e) and time measured in number of papers (right, panels b,d,f). %
(a, b) The evolution of the half-lives for 8 fields normalized with the half-lives they they had in 1970.
All half-lives measured in real time (a) show a downward trend, while when time is measured in the number of papers in the field (b), most fields show an upward trend, which indicates a slowing down in the time required to share knowledge with other fields. 
The cumulative distribution for half-lives for all subfields (SF) they are measured in (c) real time and (d) total number of papers in the system.
Panels (e) and (f) show the same distribution for journals (J).
}
\label{fig:distribution_panel}
\end{figure}

In Fig.\ref{fig:distribution_panel}b,d,f we show the half-lives that are fitted in terms of published papers. Interestingly, now the half-lives do not in general decrease as a function of the publication year as only \emph{chemistry} shows such a clearly decreasing pattern. The other fields, instead, either remain constant or show a significant increase in their half-lives over time. 
A very similar result can be seen in the half-life distributions for subfields and journals, which are shifted to higher numbers of papers as the initial time increases. That is, these results suggest that the growth in the speed of diffusion can be explained by the increase in the rate at which publications are published. 

\definecolor{tableneg}{rgb}{1.0,0.0,0.0}
\definecolor{tablepos}{rgb}{0.0,0.5,0.0}
\begin{table} %
\caption{
The half-lives in years ($t^{\frac{1}{2}}$) and the asymptotic fractions ($\beta$) for a subset of fields in 1970 and 1995 when the evolution of the diffusion process is fitted to Eq. \ref{eq1}. The relative changes from 1970 to 1995 in these values are given in the last two columns. 
}
\label{tab:fields_1970}
\footnotesize
    \begin{tabular}{ l @{\hspace{0.3cm}} c @{\hspace{0.3cm}} c @{\hspace{0.6cm}} c @{\hspace{0.3cm}}c@{\hspace{0.3cm}}c @{\hspace{0.2cm}}c}
    &  \multicolumn{2}{ c @{\hspace{0.6cm}}}{1970}&\multicolumn{2}{ c @{\hspace{0.3cm}}}{1995} &\multicolumn{2}{ c }{$\Delta$} \\
{\bf Field}  & $t^{\frac{1}{2}}$ & $\beta$ & $t^{\frac{1}{2}}$ &$\beta$ &$ t^{\frac{1}{2}}$&$ \beta$\\ \hline

    \hline
Philosophy         & 19.7  & 0.84 & 4.36 & 0.90 &{\color{tableneg}-78\%} &{\color{tablepos}+7\%}\\
Economics          & 11.0  & 0.83 & 4.20 & 0.76 & {\color{tableneg}-62\%} & {\color{tableneg}-8\%} \\
Psychology         & 8.93  & 0.72 & 3.44 & 0.67 & {\color{tableneg}-61\%} & {\color{tableneg}-7\%} \\
Linguistics        & 8.86  & 0.87 & 3.02 & 0.90 & {\color{tableneg}-66\%} & {\color{tablepos}+3\%}  \\
Chemistry          & 8.55  & 0.80 & 1.99 & 0.80 & {\color{tableneg}-77\%} & 0\%  \\
Music \& Dance     & 7.83  & 0.82 & 6.18 & 0.98 & {\color{tableneg}-21\%} & {\color{tablepos}+20\%}  \\
Gen. Humanities    & 7.25  & 0.85 & 3.43 & 0.95 & {\color{tableneg}-53\%} & {\color{tablepos}+12\%} \\
Mathematics        & 7.14  & 0.87 & 3.21 & 0.79 & {\color{tableneg}-55\%} & {\color{tableneg}-9\%} \\
Medicine           & 6.54  & 0.83 & 3.20 & 0.85 & {\color{tableneg}-51\%} & {\color{tablepos}+2\%}  \\
Sociology          & 6.34  & 0.80 & 3.72 & 0.73 & {\color{tableneg}-41\%} & {\color{tableneg}-9\%} \\
Engineering        & 4.89  & 0.82 & 2.33 & 0.79 & {\color{tableneg}-52\%} & {\color{tableneg}-4\%} \\
Law                & 4.38  & 0.92 & 7.21 & 0.80 & {\color{tablepos}+65\%} & {\color{tableneg}-13\%}\\
Social Sciences    & 4.38  & 0.73 & 2.35 & 0.59 & {\color{tableneg}-46\%} & {\color{tableneg}-19\%} \\
Physics            & 4.01  & 0.82 & 2.32 & 0.81 & {\color{tableneg}-42\%} & {\color{tableneg}-1\%} \\
Management         & 3.72  & 0.78 & 3.60 & 0.66 & {\color{tableneg}-3\% } & {\color{tableneg}-15\%} \\
Biology            & 3.43  & 0.71 & 1.69 & 0.70 & {\color{tableneg}-51\%} & {\color{tableneg}-1\%} \\
Multidisciplinary  & 1.33  & 0.59 & 1.08 & 0.59 & {\color{tableneg}-19\%} & 0\% \\
\hline
\end{tabular}
\end{table}

\section{Discussion and Conclusions}

Ever since bibliometric data of scientific publications have been available, there has been efforts to analyse such data with the goal of quantifying scientific research. The dominant framework has been to use the counts of direct citations between publications, journal, and research fields in order to quantify the relationships between them, and, for example, to rank authors \cite{10.2307/4152261}, publications  \cite{Garfield1955}, universities \cite{vanRaan2005}, and institutions \cite{Boyack2003}. As these methods work reasonably well \cite{ASI:ASI20262}, and apart from few exceptions \cite{Chen20078,PhysRevE.80.056103}, the improvements of these methods have been correcting technical flaws \cite{Seglen1997,Frey2010}. However, most of the previous research has ignored the intrinsic conceptual issue of methods based on just counting citations: they ignore the fact that scientific publications are not only based on the information created in the publications they cite, but on the whole body of literature underneath this first layer of publications.

We have introduced two simple methods to analyse how the knowledge created in a publication, or in a group of publications, might percolate through the scientific literature. This approach follows the tradition of modelling the dynamics on networks, where a real observed network is used as a substrate where the progress of a stylized model is tracked. This approach has been extensively used in network science to study epidemic spreading \cite{Pastor2015Epidemic} and social dynamics \cite{RevModPhys.81.591}. In all of these cases, the models are not expected to exactly mimic the real behaviour, but the goal is to reproduce the behaviour in the large scale with an accuracy that is enough to make at least qualitative statements about the system.  Our goal in this work was to explore this approach in knowledge spreading in citation networks. 

The first of our two measures, \textit{persistent influence}, is based on the idea of papers inheriting the knowledge of papers they cite, \textit{i.e.}, the ``shoulders'' on which they stand. As a consequence, a paper can influence later papers through chains of citations. As expected, the out-degree, in other words the direct citation count, is positively correlated with the persistent influence, but we also observe that papers that are similar in publication date and citation count can have a wide range of the persistent influence values. In our simulations we have found publications that have several orders of magnitude higher influence values than papers with similar numbers of citations. This finding suggests that there are publications for which the citation count can be a poor proxy for tracking the global cumulative influence.

We tested the hypothesis that the discrepancies in the citation counts and global persistent influence values are not meaningful but simply noise added by the global process, and to do this we used papers associated with Nobel Prizes as a manually curated corpus for presumably high influence on science. We found a significant over-performance in terms of the persistent influence by the Nobel papers when compared to the publications with similar intial citation counts and publication dates. Thus the indirect influence (i.e., the persistent influence) and direct influence (i.e., the citation count) seem to capture different aspects of the influence of a publication. Furthermore, we looked at the papers that have the greatest increase in rank while switching from the local to the global scenario, and found that these papers are often early publications in the fields that would later become hot topics of their time in the scientific world.

The second modelling approach we employed was a simple \textit{diffusion} method. We focused on analysing the rate of diffusion of knowledge across the fields, subfields, and journals. We summarized the curves describing the loss of diffusing knowledge to other fields, subfields, and journals by an exponential decay function that reaches a plateau value. Each starting time and set of seed publications can thus be described by a plateau value of the retained knowledge $\beta$ and by a typical time required $t^{1/2}$ to share half of the available knowledge. We found that $\beta$ varies heavily across disciplines, yet remaining constant in time, while the values for the half-life, $t^{1/2}$, have been steadily decreasing, suggesting an increase in the interdisciplinarity of research. However, we showed that the faster sharing of information could be explained by the increase in the rate at which publications are produced.

One crucial difference between the diffusion and persistence influence models is that in the diffusion model the total amount of knowledge created in some publication is conserved across papers citing it, whereas in the persistent influence model the knowledge is allowed to ``duplicate'' across papers. Given that there is no reason why many publications could not contain a same piece of knowledge, out of these two models the persistent influence approach is likely to provide a more realistic picture of how knowledge spreads in citation networks. The diffusion on the other hand could be considered as more of a theoretical tool to understand the structure of the citation networks.

The work done here forms a basis for future possibilities of the model-based approaches to track global knowledge spreading in citation networks. For example, more detailed look at the long-term destinations of influence starting from various sources could bring interesting results. Note also that nothing would stop one, to introduce the initial persistent influence to a group of papers (instead of a single publication) similar to the diffusion process, and repeat the type of analysis done here for fields, subfields and journals using the persistent influence model. Furthermore, one can easily reverse the tracking direction of the persistent influence model and investigate which publications, or groups of publications, in the history have influenced individual papers. One can also make the influence spreading more realistic with the cost of increasing the complexity of the model. For example, the amount of knowledge created by each publication can be added as a parameter, which will effectively work as a damping factor that will decrease the influence of very long chains of citations.

We have introduced relatively simple methods to analyse the spreading of knowledge in citation networks at a global scale, and shown that these methods can lead to significantly different results than what can be obtained by using the local approach. With more and more bibliometric data being available, we hope that our findings will encourage future work to analyse science for what it is and has always been: a cumulative process that builds over time in which the successes in scientific discoveries are built on chains of previous successes.

\section{Acknowledgements}

We used data from the Science Citation Index Expanded, Social Science Citation Index and Arts \& Humanities Citation Index, prepared by Thomson Reuters, 
Philadelphia, Pennsylvania, USA, Copyright Thomson Reuters, 2013. This research is partially supported by the European Community’s H2020 Program under the scheme  
ìINFRAIA-1-2014-2015: Research Infrastructures’, grant agreement 654024 SoBigData: Social Mining. K.K. would also like to acknowledge financial support by the Academy of Finland Research project (COSDYN) 
No. 276439 and EU HORIZON 2020 FET Open RIA project (IBSEN) No. 662725.

\setcounter{figure}{0}    
\setcounter{table}{0}    

\appendix

\section{Data Description}
\label{sec:data}

We use the data set that consists of all publications (publications and reviews) written in English from the year 1898 till the end of 2013, included in the database of the Thomson Reuters (TR) and now Clarivate Analytics' Web of Science. The data set contains a journal assignment for most publications and most journals are further assigned to one or more subfields. We filter out publications and journals for which this information is not available, which leaves us around 35 million publications in around 15 thousand scientific journals. We further map the subfields of the publications into major scientific fields as in Ref.\cite{PDBP}.

We use the set of citations between the filtered publications to construct a network where there is a link from citing publication to the cited publication. We use the publication time information to remove links where the date of the cited publication is equal or later than the date of the citing publication. In total we remove $1.7 \%$ of the links this way, and we are left with a citation network without any cycles, \emph{i.e.}, a directed acyclic graph. To avoid boundary effects for the latest publications, for which most publications citing them are not in our data set, we only consider the nodes in the citation network until the year 2008. Previous literature \cite{ASI:ASI20744} shows that the typical life cycle of a publication in terms of citation is completed within ~5 years from date of publication. Because the data used here ends in 2013, limiting our attention to publications published until 2008 minimizes the boundary effects originating from the missing data on future publications.

\section{Alternative diffusion methods}
\label{sec:alt_diffusion}

We also tested assigning initial values asymmetrically by looking at how many citations each paper has received in the first 5 years (plus one, to take care of citationless papers): $D_{i} = \frac{1+C_{i}(t<5)}{\sum_{j} (1 + C_{j}(t<5) }$. This count acts as a proxy of how successful the paper has been in general, but this alternative initialization strategy resulted in qualitatively similar results to the more simple strategy and we do not show these results here. 

Similar to the initialization of the diffusion process, we tried out two different ways of selecting the probabilities that the random walker uses to follow citations to the future. In the simple case the walker jumps from the cited publication to each citing publication with the same probability, and in the other case the random walker preferentially chooses publications that receive more citations in the coming five years $C_{i}(t<5)$. This is equivalent to setting the weight $w_{ij}$ of each link from paper $i$ to a citing paper $j$ to one of the two values:

\begin{equation} 
\begin{split}
w_{ij} &= \frac{1}{\sum_{l \in \Gamma_{i}^{out}} 1} = \frac{1}{k_i^{out}}  \text{\,\,or\,\,} \\   
w_{ij} &= \frac{c_{j}(t<5)}{\sum_{l \in \Gamma_{i}^{out}} c_{l}(t<5)}.
\label{eq:diffusion-weights}
\end{split}
\end{equation}  

The results for both processes are similar and here we only show them for the simpler process. For technical details of how the process was made computationally tractable and implemented see Appendix \ref{sec:computations}.

\section{Computational considerations}
\label{sec:computations}

In order to implement both the diffusion and influence algorithms we had to organize the citation network in Directed Acyclic Graphs (DAGs), as mentioned in the Data Description section. After this it was necessary to order the nodes in \textit{topological order}, which guarantees that for every directed edge connecting paper $i$ (the citing paper and $j$ (the cited paper), $j$ comes before $i$. Such ordering should, in principle, correspond with the time stamp of the publication. However, for older papers the topological ordering of the publications could not be resolved using publication dates, as only year information was available. Therefore, we took advantage of the fact that papers published in earlier years are bound to have a lower rank in the order. Thus we built yearly citation networks and ordered them topologically, starting from the oldest one. We then proceeded to arrange the nodes topologically within the year network, building the overall topological order adding one layer of publications at a time. Once a topological ordering of the nodes was created, we built a topological ordering for the edges, sorting them by topological order of the cited paper. When looping through the topologically sorted edge-list, this ordering guarantees that each paper has collected all the value/influence upstream before pushing its own forward, and that all value/influence contained by the paper is pushed forward to the citing papers at once. Further, this ordering allowed us to loop through the edge-list only once and to control the start and end of the pushing process by checking for each edge that the topological ordering values of each paper lie within the year bounds.

In order to implement the diffusion process, we have chosen as seeds the set of all papers being published in the same field in the same year $y_{start}$. By doing this we are able to select a very coherent set of papers both in terms of subject and time. Once the system has been initialized, the next step is to choose a final year $y_{end}$ as the year in at which we will stop pushing values forward. Hence we loop through the node-list of all scientific papers in our dataset published between $y_{start}$ and $y_{end} -1$  arranged in topological order to initialize the weights for each node. We consider only links to neighbours that point to papers published before $y_{end}$. 

\begin{figure}[htpb!]
\centering
\includegraphics[width =.24\columnwidth]{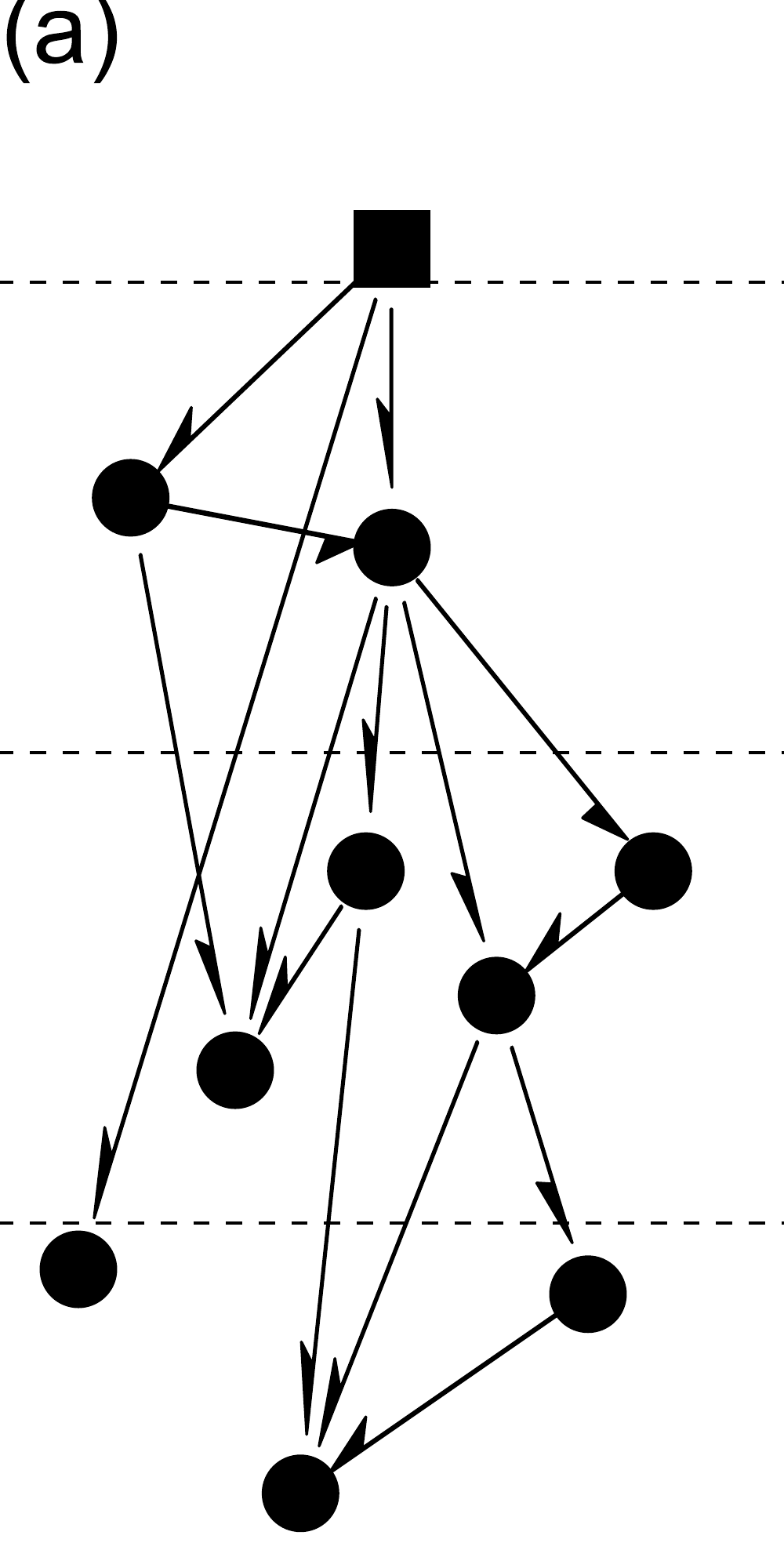}
\includegraphics[width =.24\columnwidth]{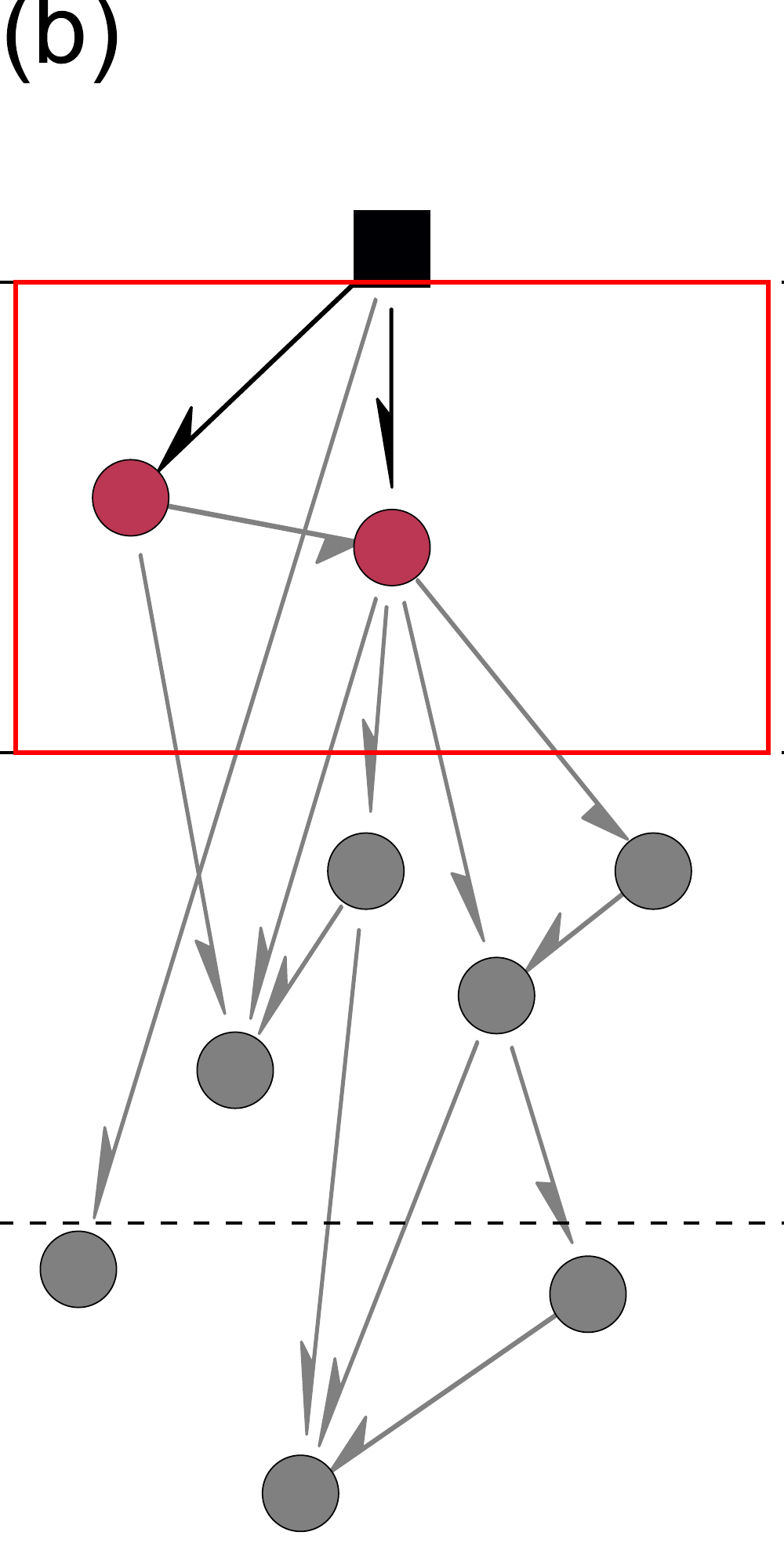}
\includegraphics[width =.24\columnwidth]{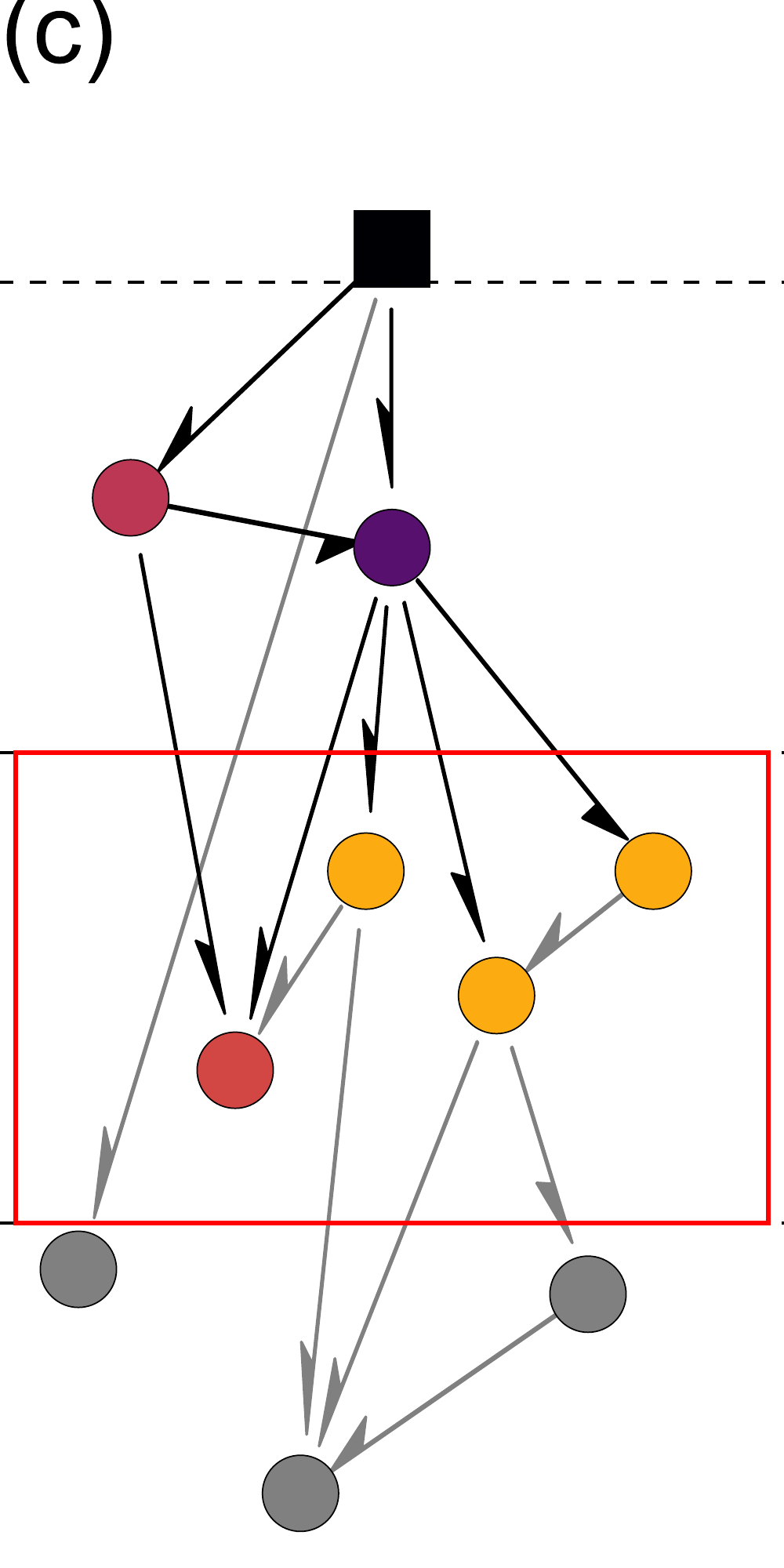}
\includegraphics[width =.24\columnwidth]{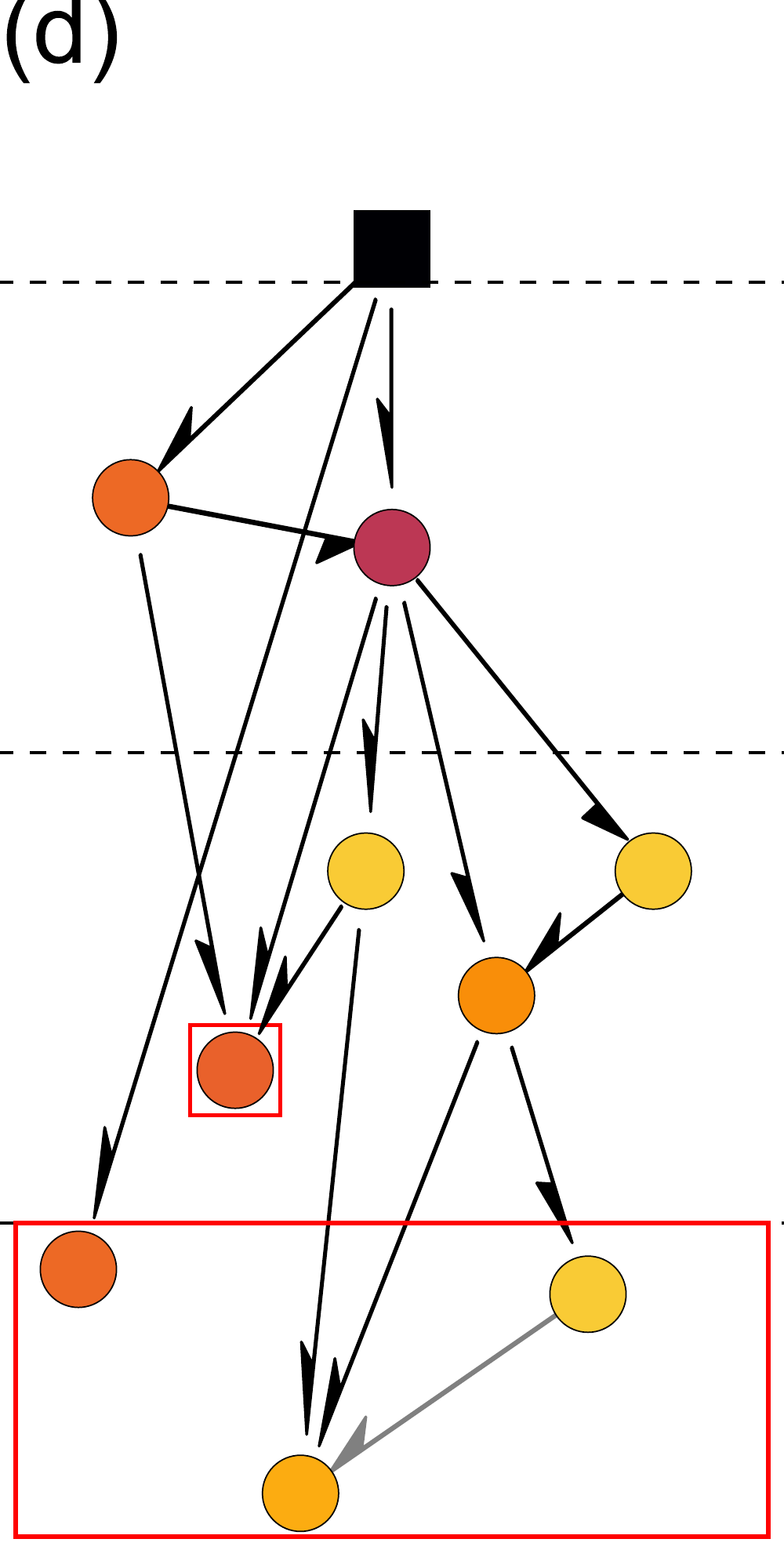}
\caption{An example of the diffusion process and the collection of data. 
Panel a shows the original DAG, while panels b,c,d show the result of the diffusion process for each year (delimited by the dotted line). The sum of the values in each red box sums up to one.  Note how in the last panel a sink exists and is included as being part of the set of papers for which to sum the values.}
\label{fig:schematic_diffusion}
\end{figure}

Once the weights for the diffusion process have been initialized we can push the value of each node by looping again through the nodelist in by topological order, which guarantees that no value is ever pushed from a node before the same node has collected all the previous values available. The pushing starts from $y_{start}$ and stops in $y_{end} -1$ but spreads to papers published all the way to $y_{end}$, without pushing any value within $y_{end}$ . This means that we consider as leaf nodes of the system only the first papers to receive value in the final year, as receiving citations in the first year is somewhat hard to obtain (it heavily depends on the month of publication) and one single citation might steal all its value from another paper. In case of the existence of a \textit{sink} (i.e. an uncited node) in previous years, we also consider it being part of the final year. However, only a small fraction of nodes remain uncited until 2008, with most papers being only temporary sinks. This means that the almost all of the value is pushed along, with little risk of it being trapped in sinks along the way.

After the pushing has ended we can collect all the values that are left not pushed in the system. Since the pushing has been carried out by following the topological order of the whole graph, this is simply accomplished by not storing permanently the value of any node that appears in the first column of the edge-list as by definition they will necessarily get rid of all their values. Also, by construction the sum of the values of all the leaves equal one. It is important to notice that in order to collect the data between say 1990 and 2008 one needs to repeat the pushing process for each $y_{end}$ between those years, since the network initialized is different each time. This means that when we collect the data in a certain year, we do not consider what happened in the future (except the 5 year citation proxy). If we were to collect the data in middle years while pushing the values directly to the last year, we would be including links to recent papers that would steal value from the middle years, thus altering the renormalization factor. The data collection, like the data initialization, can be done on paper, journal, subfield and field level.

In Fig.~\ref{fig:schematic_diffusion} we show an example of the diffusion process and the data collection. We can see that the method explained the previous paragraph causes the intermediate values (e.g. the values of the first time interval in Fig.~\ref{fig:schematic_diffusion}) to be modified when the process is run for more recent years. This is because, new publications and their links to older links cause a change in the link weights computed using Eq.~\ref{fig:schematic_diffusion}.

\section{Highest Cited Papers}

\begin{table}[H]\footnotesize
\caption{Publications with highest citation rank for each year.}
\label{tab:highest_citations}
  \tiny
    \begin{tabular}{ c  c c|p{6.5cm}}   
    $R_c$ & $R_I$&  Year & Title \\ \hline
 1 & 1 & 1970 & Cleavage Of Structural Proteins During Assembly Of Head Of Bacteriophage-T4\\   
 1 & 63 & 1971 & The Assessment And Analysis Of Handedness: The Edinburgh Inventory\\   
 1 & 1 & 1972 & Regression Models And Life-Tables\\   
1 & 19 & 1973 & Relationship Between Inhibition Constant (K1) And Concentration Of Inhibitor Which Causes 50 Per Cent Inhibition (I50) Of An Enzymatic-Reaction\\   
1 & 1 & 1974 & Film Detection Method For Tritium-Labeled Proteins And Nucleic-Acids In Polyacrylamide Gels\\   
 1 & 1 & 1975 & Detection Of Specific Sequences Among Dna Fragments Separated By Gel-Electrophoresis\\   
 1 & 1 & 1976 & Rapid And Sensitive Method For Quantitation Of Microgram Quantities Of Protein Utilizing Principle Of Protein-Dye Binding\\   
 1 & 1 & 1977 & Dna Sequencing With Chain-Terminating Inhibitors\\   
1 & 11 & 1978 & Rapid Chromatographic Technique For Preparative Separations With Moderate Resolution\\   
 1 & 1 & 1979 & Electrophoretic Transfer Of Proteins From Polyacrylamide Gels To Nitrocellulose Sheets - Procedure And Some Applications\\   
 1 & 6 & 1980 & Ligand - A Versatile Computerized Approach For Characterization Of Ligand-Binding Systems\\   
 1 & 1 & 1981 & Improved Patch-Clamp Techniques For High-Resolution Current Recording Fromcells And Cell-Free Membrane Patches \\   
 1 &1 &  1982 & A Simple Method For Displaying The Hydropathic Character Of A Protein\\   
1 & 1 & 1983 & A Technique For Radiolabeling Dna Restriction Endonuclease Fragments To High Specific Activity\\   
1 & 2 & 1984 & A Comprehensive Set Of Sequence-Analysis Programs For The Vax\\   
 1 & 4 & 1985 & A New Generation Of Ca-2+ Indicators With Greatly Improved Fluorescence Properties\\   
 1 & 3 & 1986 & Statistical Methods For Assessing Agreement Between Two Methods Of Clinical Measurement\\   
 1 & 1 & 1987 & Single-Step Method Of Rna Isolation By Acid Guanidinium Thiocyanate Phenolchloroform Extraction\\   
 1 & 8 & 1988 & Development Of The Colle-Salvetti Correlation-Energy Formula Into A Functional Of The Electron-Density\\   
 1 & 32 & 1989 & Gaussian-Basis Sets For Use In Correlated Molecular Calculations .1. The Atoms Boron Through Neon And Hydrogen\\   
 1 & 2 & 1990 & Basic Local Alignment Search Tool\\   
 1 & 2 & 1991 & Molscript - A Program To Produce Both Detailed And Schematic Plots Of Protein Structures\\   
 1 & 5 & 1992 & The Mos 36-Item Short-Form Health Survey (Sf-36) .1. Conceptual-Framework And Item Selection\\   
 1 & 1 & 1993 & Density-Functional Thermochemistry .3. The Role Of Exact Exchange\\   
 1 & 1 & 1994 & Clustal-W - Improving The Sensitivity Of Progressive Multiple Sequence Alignment Through Sequence Weighting, Position-Specific Gap Penalties And Weight Matrix Choice\\   
 1 & 4 & 1995 & Genepop (Version-1.2) - Population-Genetics Software For Exact Tests And Ecumenicism\\   
 1 & 2 & 1996 & Generalized Gradient Approximation Made Simple\\   
 1 & 1 & 1997 & Gapped Blast \& Psi-Blast: A New Generation Of Protein Database Search Programs\\   
 1 & 2 & 1998 & Crystallography And Nmr System: A New Software Suite For Macromolecular Structure Determination\\   
 1 & 2 & 1999 & Mechanisms Of Disease - Atherosclerosis - An Inflammatory Disease\\   \hline

     \end{tabular}
\end{table}

\section*{References}

\bibliography{main}

\clearpage

\end{document}